\documentclass[10pt,twocolumn,conference]{IEEEtran}
\makeatletter
\IEEEoverridecommandlockouts
\def\ps@headings{%
\def\@oddhead{\mbox{}\scriptsize\rightmark \hfil \thepage}%
\def\@evenhead{\scriptsize\thepage \hfil \leftmark\mbox{}}%
\def\@oddfoot{}%
\def\@evenfoot{}}
\makeatother
\usepackage[ruled,vlined]{algorithm2e}
\providecommand{\SetAlgoLined}{\SetLine} 
\providecommand{\DontPrintSemicolon}{\dontprintsemicolon} 
\usepackage{soul}
\usepackage{graphicx,cite}
\usepackage{epsfig}
\usepackage[cmex10]{amsmath}
\usepackage{amssymb}
\usepackage{array}
\usepackage{fixltx2e}
\usepackage{amsfonts}
\usepackage{lineno}
\usepackage{setspace,subfig,color,amsmath,epsfig,amssymb,booktabs} %
\usepackage{verbatim}

\newtheorem{theorem}{Theorem}[section]
\newtheorem{lemma}[theorem]{Lemma}

\makeatletter
\renewcommand\paragraph{\@startsection{paragraph}{4}{\z@}%
    {1.5ex plus .2ex minus .3ex}%
            {-0em}%
                        {\normalsize\bf}}

\newtheorem{definition}{Definition}

\begin{document}
\title{A Robust Relay Placement Framework for 60GHz mmWave Wireless Personal Area Networks}

\author{
Guanbo Zheng,
Cunqing Hua,
Rong Zheng,
and Qixin Wang\IEEEauthorrefmark{1}
\thanks{
\IEEEauthorrefmark{1}Guanbo Zheng and Rong Zheng are with Department of Electrical and Computer Engineering and Department of Computer Science, University of Houston, TX, 77004 USA;
Cunqing Hua is with the School of Information Security Engineering, Shanghai Jiao Tong University, Shanghai, 200240 China;
Qixin Wang is with the Department of Computing, Hong Kong Polytechnic University, Hong Kong, China
E-mail: {\it \{gzheng3, rzheng\}@uh.edu, cqhua@sjtu.edu.cn, csqwang@comp.polyu.edu.hk }
}
}
\maketitle

\begin{abstract}
Multimedia streaming applications with stringent QoS requirements in 60GHz
mmWave wireless personal area networks (WPANs) demand high rate and low
latency data transfer as well as low service disruption. In this paper, we
consider the problem of robust relay placement in 60GHz WPANs. Relays
forward traffic from transmitter devices to receiver devices facilitating i) the
primary communication path for non-line-of-sight (NLOS) transceiver pairs, and
ii) secondary (backup) communication path for line-of-sight (LOS) transceiver pairs.
We formulate the \emph{robust minimum relay placement}
problem and the \emph{robust maximum utility relay placement}
problem with the objective to minimize the number of relays
deployed and maximize the network utility, respectively. Efficient algorithms
are developed to solve both problems and have been shown to incur less service
disruption in presence of moving subjects that may block the LOS paths in the
environment.
\end{abstract}

\section{Introduction}
\label{sec:intro}

The \emph{millimeter wave} (mmWave) band has attracted considerable commercial
interests due to the advance in low-cost mmWave radio frequency integrated
circuit design.  The mmWave band provides 7GHz unlicensed spectrum resource at
the center frequency of 60GHz (57-64GHz in North America, and 59-66GHz in Europe and Japan),
which would enable many high data rate applications
like high definition streaming multimedia, high-speed kiosk data transfer and
point-to-point terminal communication in data center~\cite{anderson04,
daniels10, flyway2011}, etc.

In contrast to many existing RF technologies such as 2.4GHz WiFi radios, mmWave
radio has several unique physical characteristics. First, the propagation and
attenuation loss are much more severe in 60GHz band. It is shown that the free
space path loss in 60GHz is more than 20dB larger than that in 5GHz. The
oxygen absorption loss is as high as 5-30dB/km. Furthermore, the penetration
loss is also much higher through typical building
materials~\cite{anderson04}. As a result, \emph{line-of-sight}
(LOS) path is the predominant path for signal transmission, while signals along the
second-order and higher-order reflection paths are significantly attenuated and often
negligible. Second, to combat such severe signal degradation, directional
antenna technology is essential in mmWave devices. By using directional antennae
on both the transmitter and the receiver sides, mmWave radios can obtain
significant gain in the received signal strength, while incurring negligible
interference to/from other mmWave devices~\cite{yiu09, singh09, singh10}.
In this paper, we consider an mmWave wireless personal area network (WPANs) equipped with directional antennae on all devices.

In addition to high bandwidth demands, multimedia applications in 60GHz WPANs
also have stringent requirements on service disruption (defined as the
frequency and duration of time that network connectivity is not available),
which can occur due to changes in channel conditions such as LOS link blockage
by moving subjects in the space. This motivates us to employ relays for two
purposes.  First, relays can be used to forward traffic from transmitters to
receivers that do not have LOS connectivity.  Second, relays can provide a
secondary (2-hop) path in case of blockage on the primary (direct) path.

Generally, there are two types of relays proposed for mmWave in the literature,
active relay~\cite{singh07,yiu10,Jian-twc11} and passive relay~\cite{Alberto,Kenneth,xia2012}.
A passive relay (also known as reflector) reflects the mmWave radiation from transmitter to receiver.
It can be as simple as a flat metal plate that does not require any power source.
However, it introduces losses due to reflection, as well as the additional path loss as a result of longer propagation path.
In contrast, an active relay is an active mmWave transceiver with beamforming capabilities.
It can amplify and forward the mmWave signal from the transmitter to any intended direction, at the cost of higher complexity and maintenance.
In this paper, we consider {\it active} relays for the ease of control of reflection directions
and signal boost.

In this paper, two relay placement problems in 60GHz mmWave WPANs are
investigated: \emph{robust minimum relay placement} (RMRP) that attempts to
find the minimum number of relays and their best placements from a set of
candidate locations with bandwidth and robustness constraints, and
\emph{robust maximum utility relay placement} (RMURP) that aims to maximize
network utility given a fixed number of relays.
Two vertex-disjoint (except for the endpoints) paths (one called the primary path, and
the other the secondary path) are provisioned between each pair of transmitter
and receiver. Consequently, \emph{seamless} switching to the secondary path is
facilitated in event of channel degradation or blockage on the primary path
avoiding service disruption.  Robustness is characterized by the \emph{D-norm}
uncertainty model, which models tolerance to concurrent (worst-case) failures
of a subset of primary paths\cite{yang091}. Using linear relaxation and the
duality theorem of linear programming, RMRP and RMURP are transformed
to the mixed integer linear programming (MILP) and mixed integer non-linear
programming (MINLP) problems, respectively. Two algorithms, {\it bisector search} and {\it
Generalized Benders Decomposition},  are proposed for RMURP and are shown to
have near optimal and optimal performance. Extensive simulations demonstrate the
fault tolerance of the proposed relay placement algorithms in significantly
reducing the probability of service disruption.

The rest of this paper is organized as follows. The related literature work is
analyzed in Section~\ref{sec:related}.  In Section~\ref{sec:problem_form}, we
introduce the network model and the problem statements for robust relay
placement.  The RMRP formulation is presented in Section~\ref{sec:rmrp}, while
RMURP is formulated in a similar way in Section~\ref{sec:rmurp}.  Furthermore,
we also discuss two proposed algorithms for RMURP in Section~\ref{sec:rmurp}.
Performance evaluation is presented in Section~\ref{sec:performance}.  Finally,
we conclude this paper in Section~\ref{sec:conclusion}.

\section{Related Work}
\label{sec:related}

Significant prior literature have been produced on different aspects of 60GHz
radios, from CMOS circuit design to network protocol development.  In this
section, we summarize prior work on MAC design in mmWave WPANs.

A spatial time-division multiple access (STDMA) scheme was proposed for a
realistic multi-Gbps mmWave WPAN in \cite{sum-gc09}.  With the help of a
heuristic scheduling algorithm, it is able to achieve significant
throughput enhancement as much as 100\% compared to conventional TDMA schedules.
In \cite{cai-wcnc07}, Cai {\it et al.} presented an efficient resource management
framework based on the unique physical characteristics in a MC-DS-CDMA based
mmWave networks. The authors also conducted extensive analysis of spatial
multiplexing capacity in mmWave WPANs with directional antennae in
\cite{cai-gc07, cai-twc10}.  In~\cite{singh09}, Madhow {\it et al.} conducted a
probabilistic analysis of the interference in an mmWave network, as the result
of uncoordinated transmission. It is concluded that an mmWave link can be
abstracted as a ``pseudo-wired link'' with negligible interference when the
beam width is 20 degree. Similar observations are made
in~\cite{singh-jsac09, yiu09}. Therefore, the primary interference at the
transmitter or receiver devices is the predominant source of contention.  In
\cite{gong-wcnc10, gong-gc10}, to address the deafness problem induced by
directionality, Gong {\it et al.} proposed a new directional CSMA/CA protocol
for IEEE 802.15.3c 60GHz WPANs. With virtual carrier sensing, the central
coordinator can distribute the network allocation vector (NAV) information,
to avoid collisions among the devices occupying the same channel.
The author also extended the work to a multiple-user scenario in
\cite{gong-icc11}.  A distributed scheduling protocol is proposed by
coordinating mmWave mesh nodes in~\cite{singh10}, and can achieve high
resource utilization with time division multiplexing (TDM). However, none of
the above work model or address relay placement problems in mmWave WPANs
with directional antennae.

There are also some existing work on repeater selection and relay operation
scheduling in mmWave WPANs.  Repeater selection was investigated
in~\cite{yiu10}, with the objective to maximize data rate for each transmitter
and receiver pair by determining the best link allocation.
In~\cite{wang10_vtc}, Lan {\it et al.} explored time slot scheduling for relay operations in the scenario
of directional antenna on mmWave devices and formulated the throughput
maximization problem as an integer programming problem. However, both
schemes do not consider robustness in presence of uncertain link blockage.

Our work is also related to multihop routing in wireless networks with directional antenna~\cite{Zhou2009,Partha2010,di2010} with two key differences.
First, relays in our work are dedicated devices that do not generate or receive application layer packets.
Second, we allow at most 2-hop paths between any mmWave transmitter-receiver pair,
considering the fact that mmWave WPANs are deployed in small indoor environment with stringent QoS requirement.

To the best of our knowledge, we are the first to explore robust relay
placement in 60GHz WPANs. Some preliminary results are presented in
\cite{zheng-rtas12}.

\section{Network Model and Problem Statement}
\label{sec:problem_form}

Consider an mmWave network consisting of a set of $L$ \emph{logical mmWave
links} (simplified as \emph{logical links}), each link $i \in L$ is
associated with a source device (transmitter) $s_i$, a destination
device (receiver) $d_i$, and a traffic demand $r_i$ bps. \emph{mmWave
relay devices} (simplified as \emph{relays}) equipped with steerable
antennae can relay data between the transmitters and receivers. The relays
can be placed at a set of $K$ candidate locations. We further consider a set of $O$
obstacles in the environment with known locations.

\subsection{Geometric Model for Link Connectivity}
\label{sec:preliminary}

In this section, we introduce a geometric model to characterize link
connectivity in 60GHz mmWave WPANs. LOS transmissions are feasible between a
transmitter and a receiver if and only if the direct path between them is
unobstructed and their distance is less than a threshold $d$.

To bound the end-to-end latency, at most two-hop paths (involving one
relay) are allowed between any transmitter-receiver pair.
The proposed solutions can be extended to cases where longer paths are allows.
Thus, the connectivity of any logical link is determined by the visibility
regions of its end points defined as follows:
\begin{definition}({\bf Visibility region})
Given a 2-D plane of interest, any two points $(a, b)$ are visible to each
other if the line segment between them does not intersect with any obstacles
and the distance of $a$ and $b$ is no more than $d$. The \emph{visibility
region} $V(a)$ of a point $a$ in the plane is the bounded shape consisting of
all unobstructed points no more than distance $d$ from $a$.
\end{definition}

Consider transmitter and receiver $a$ and $b$, the (binary) connectivity of
logical link $(a, b)$ is thus characterized by:
\begin{equation}
\label{overlap_region}
\lambda(a, b) = \begin{cases} 1, & \text{iff } V(a) \cap V(b) \neq \varnothing
\\ 0, & \text{otherwise } \end{cases} ,
%
\end{equation}
If $\lambda(a, b)=1$, logical link $(a, b)$ is \emph{feasible} (directly or via a
relay in $a$ and $b$'s overlapping visibility region); otherwise, it is
\emph{infeasible}. For the rest of the paper, we only consider the set of feasible logical links given by:
\begin{equation} \label{feasible_linkset}
\Omega = \{ i \mid \lambda(s_i, d_i) = 1, \forall i \in L \},
\end{equation}
\noindent where $s_i$, $d_i$ are the transmitter and receiver  of the $i$-th
logical link respectively.

\begin{figure}[t!]
\centering
\includegraphics[width=2.5in]{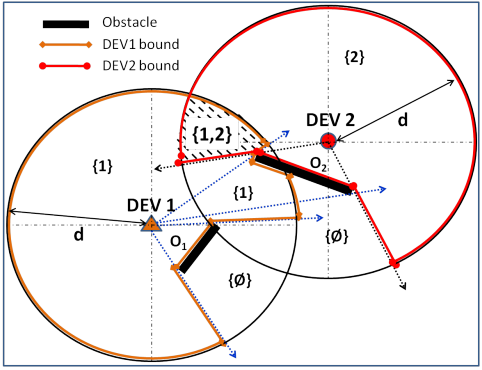}\\
\caption{Visibility region and overlapped visibility region.}
\label{fig:visibility_fig}
\end{figure}

Fig.~\ref{fig:visibility_fig} illustrates the notion of visibility region.
There are two mmWave end devices, DEV1 and DEV2.  The shadowed area corresponds
to the overlapped visibility region between DEV1 and DEV2, which is the
candidate region for relay placement for DEV1 and DEV2.  The existence of
overlapped visibility region is the necessary and sufficient condition of the
connectivity of logical links, in both LOS and \emph{non-LOS} (NLOS) cases.

\newcommand{\MK}{\mathcal{K}}

The relationship between feasible logical links and candidate relay
locations can be modeled as an undirected bi-partite graph $G(\Omega, \MK, E)$,
where $\MK = {1, \ldots, K}$ is the set of candidate relay locations.  An edge $e=(l, k)$
exists between a logical link $l \in \Omega$ and a candidate relay location
$k \in \MK$ iff $k$ is in $V(s_l)\cap V(d_l)$, where $s_l$ and $d_l$ are the
end points of $l$. Thus, the set of feasible logical links that can use a
relay placed at the candidate location $k$ as relay is given by:
\begin{equation} \label{feasible_linkset_k}
\Omega_k = \{ i \mid k \in V(s_i) \cap V(d_i), \forall i \in \Omega \}, \forall
k \in \MK. \\
\end{equation}

\subsection{Needs for robustness}
\label{sec:robust}

Relays serve two purposes: {\it i)} providing the primary communication
path for NLOS logical links; and {\it ii)} providing secondary (backup)
communication path for LOS or NLOS logical links. Provisioning of secondary
paths reduces service disruption when the primary path is
obstructed.

\begin{figure*}[tbh!]
\centering
\begin{tabular}{ccp{5.4cm}}
\includegraphics[width=2.45in]{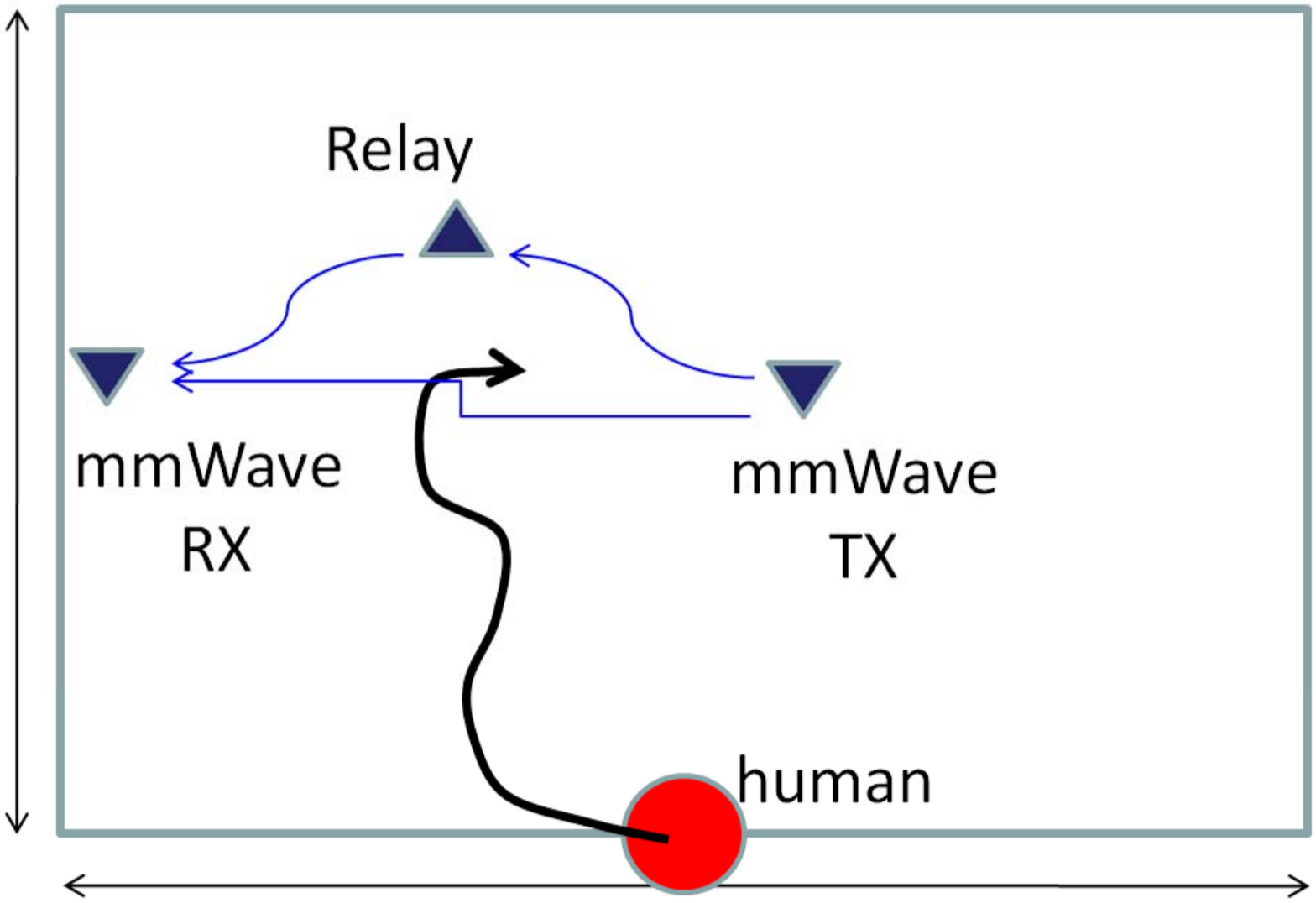} &
\includegraphics[width=2.2in]{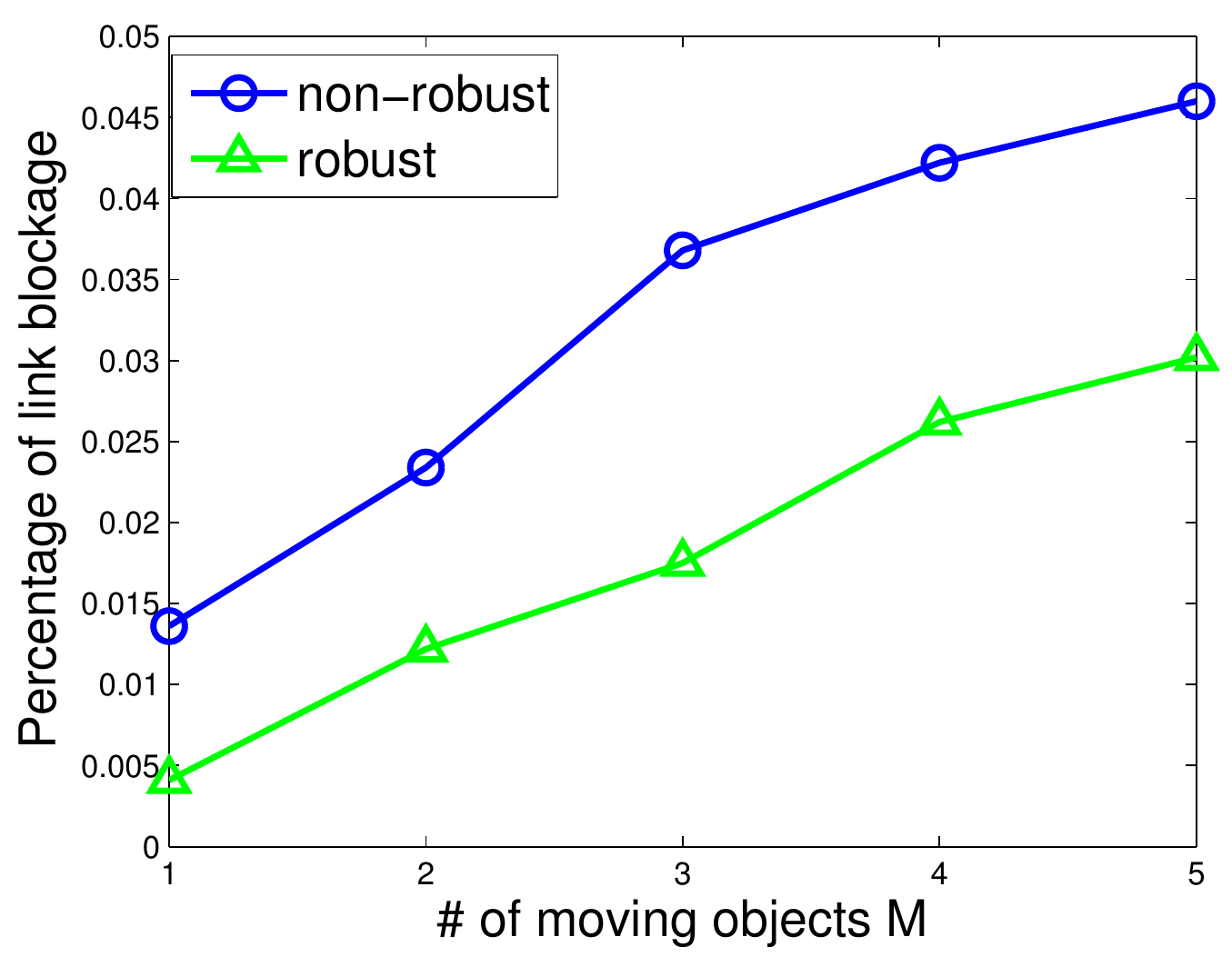} &
\includegraphics[width=2.05in]{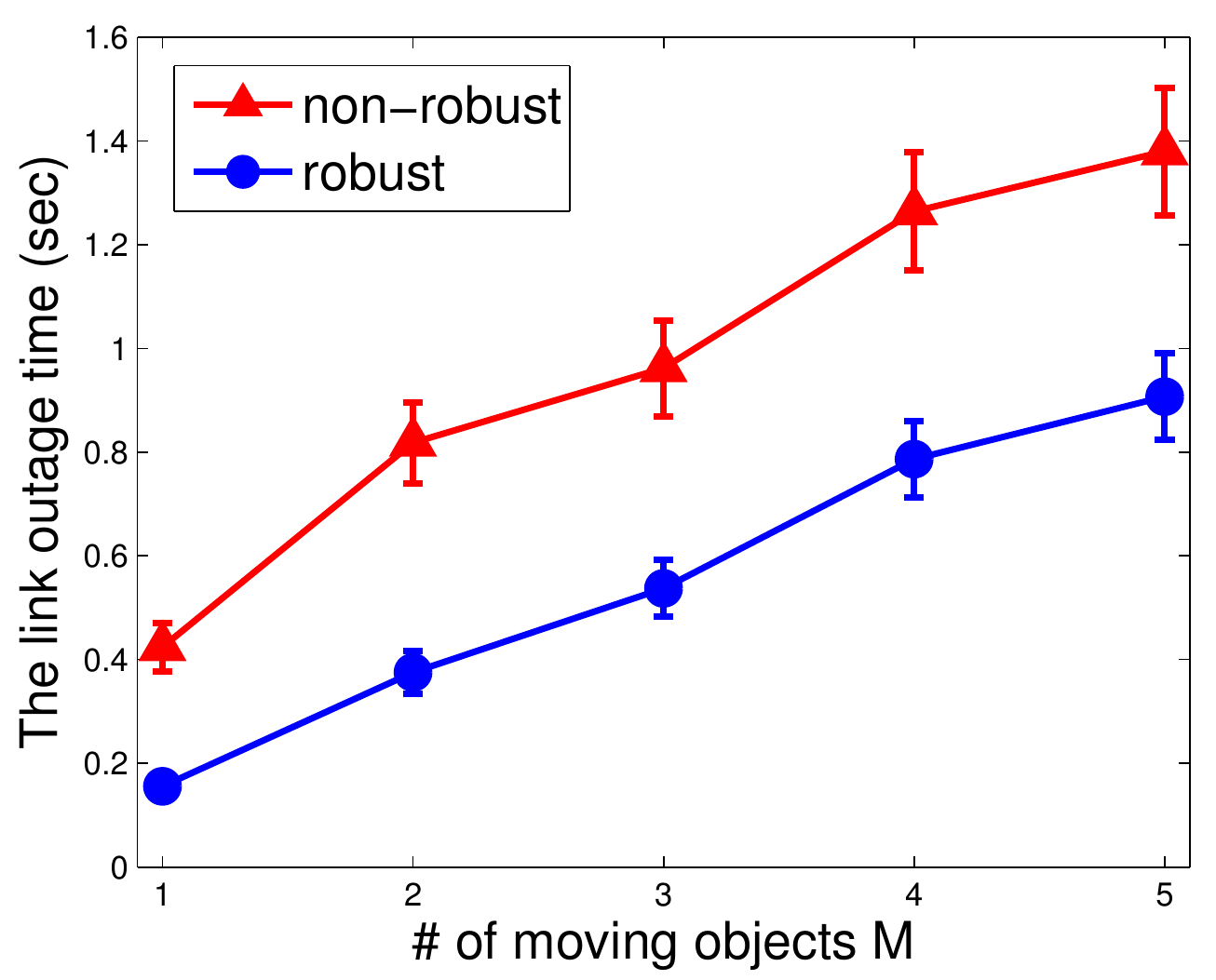} \\

(a) Simulation setup & (b) Percentage of link blockage & (c) Mean blockage duration with 90\% confidence interval\\
%
%
%
%
\end{tabular}
\caption{Link blockage with and without relay due to moving human subjects
in a 10m $\times$ 10m room with a fixed mmWave TX/RX and a dedicated relay,
$M$ human subjects moving randomly inside.}
\label{fig:profailure}
\end{figure*}

To see the impact of secondary paths, we conduct a simple simulation study.
Consider a home-network environment in Fig.~\ref{fig:profailure}(a), where
there is a LOS logical link $l$ and a dedicated relay at a fixed location.
In the robust setting,  one relay is used to provide a secondary
communication path for $l$.  Inside the room, there are $M$ moving human
subjects modeled as a circle with a radius of 0.3 meters. We adopt a classic random
walk model~\cite{Bai_mobility}, where in each step, a person moves 0.3 meters
with the direction randomly chosen from the set $\{-90^o, -45^o, 0^o, 45^o,
90^o\}$. Without relays, the communication between TX and RX is disrupted
when a person blocks the direct LOS path. With relays, an outage occurs only
when both the primary (LOS) path and the secondary path (via the relay) are
blocked.  Fig.~\ref{fig:profailure}(b) and Fig.~\ref{fig:profailure}(c) show
the percentage of link blockage and the mean blockage duration with 90\% confidence interval
versus different number of moving human subjects, respectively.

As shown in Fig.~\ref{fig:profailure} (b)(c), when the number of moving subjects
increases, the percentage of link blockage and blockage duration increase with and without the
secondary path. However, the use of backup path reduces both the blocking
probability and the duration of each outage. This translates to better quality
of service (QoS) at the application layer.

\subsection{Problem Statement}
Relay placement concerns with the selection of relays among a finite
set of candidate locations to optimize for certain network utilities. We
consider two variations of the problem.
\begin{definition} ({\em Robust Minimum Relay Placement (RMRP) problem})
Given an mmWave network with a set of feasible logical links with fixed
traffic demands, find the minimum number of relays and their locations among
candidate location set $\MK$ that satisfy connectivity, bandwidth and
robustness constraints.
\end{definition}

\begin{definition} ({\em Robust Maximum Utility Relay Placement (RMURP)
problem})
Given an mmWave network with a set of feasible logical links with base
traffic demands, find the placement of at most $m$ relays among candidate location set
$\MK$ such that the ratio of the achievable rates over the base rate is
maximized subject to robustness constraints.
\end{definition}

We restrict the relay of data to relays only. In the robust formulation, for
each feasible logical link, two vertex-disjoint (except for the endpoints)
communication paths are provisioned, one as \emph{primary path}, and the other
as \emph{secondary path}.  Both the primary and secondary paths between mmWave
transmitters and receivers cannot be more than 2-hops.  If a relay serves
as relay for more than one logical links, \emph{time division medium access}
(TDMA) scheduling is adopted.  The interference among concurrent transmissions
is assumed to negligible due to the high directionality of mmWave
communications, which is consistent with the measurement results reported in
\cite{singh10}. Therefore, the main sources of contention arise from the
half-duplex constraint and multiplexing at the relay nodes.
%
\section{Robust Minimum Relay Placement (RMRP)}
\label{sec:rmrp}

In this section, we present the analytical form of the RMRP problem.  The
following notations are used:
\begin{itemize}
\item{\bf Primary indicator}: $x_{ik} = 1$ if relay $k$ is selected by logical
link $i$ as its primary path relay; otherwise, $x_{ik} = 0$;
\item{\bf Secondary indicator}: $y_{ik} = 1$ if relay $k$ is selected by
logical link $i$ as its secondary path relay; otherwise, $y_{ik} = 0$;
\item{\bf Selection indicator}: $z_k= 1$ if relay $k$ is selected by at
least one logical link; otherwise, $z_k= 0$.
\item{\bf NLOS indicator}: $\eta_i = 1$ if logical link $i$ does not have a LOS
path; otherwise, $\eta_i = 0$.
\end{itemize}

If a logical link $i$ has a LOS path, no relay is needed for the primary
path; otherwise, one relay should be selected for the primary path, which
should satisfy the following condition:
\begin{equation} \label{eq5_x}
\sum_{k=1}^{K} x_{ik}=\eta_i,  \forall i \in \Omega.
\end{equation}

On the other hand, at least one relay is needed to facilitate the secondary
path, that is,
\begin{equation} \label{eq5_y}
\sum_{k=1}^{K} y_{ik}=1, \forall i \in \Omega.
\end{equation}

In addition, a relay cannot be used for the primary path and the secondary path
simultaneously. Therefore, we have:
\begin{equation} \label{unique_ref}
x_{ik} + y_{ik} \le 1, \forall i \in \Omega_k, \forall k.
\end{equation}

As mentioned before, for simplicity, we assume each relay has only one
half-duplex transceiver. Therefore, the transmission time of relaying a unit data of logical link
$i$ via relay $k$ is
\begin{equation} \label{tx_time}
\tau_{ik} = \frac{1}{R_{s_i, k}} + \frac{1}{R_{k, d_i}},
\end{equation}
\noindent where $R_{s_i, k}$, $R_{k, d_i}$ are the mmWave data bandwidth
between the source and the relay, and between the relay and the
destination, respectively.  For the AWGN channel, they can be modeled as:
\begin{equation} \label{rate_sr}
R_{s_i, k} = \begin{cases}
& W\log \left[ 1 + \frac{P_t G_t G_r}{P_n D(s_i, k)^\gamma }\right] \mbox{
(when $D(s_i,k) \leq \Theta$)} \\
& 0 \mbox{   (when $D(k,d_i) > \Theta$),}
\end{cases}
\end{equation}

\noindent and
\begin{equation} \label{rate_rd}
R_{k, d_i} = \begin{cases}
& W\log \left[ 1 + \frac{P_t G_t G_r}{P_n D(k, d_i)^\gamma } \right] \mbox{
(when $D(k,d_i) \leq \Theta$)} \\
& 0 \mbox{   (when $D(k, d_i) > \Theta$),}
\end{cases}
\end{equation}
\noindent respectively, where $W$ is the channel bandwidth in Hz, $P_t$ is the
transmission power, $P_n$ is the noise floor level, $G_t$, $G_r$ are
transmitter and receiver antenna gains, $\gamma$ is the large-scale path loss
index, $D(a, b)$ is the distance from $a$ to $b$, and $\Theta$ is a constant
threshold on communications radius.

For a relay $k$, the TDMA scheduling for the associated logical links should
satisfy:
\begin{equation} \label{scheduling}
\sum_{i \in \Omega_k} \eta_i x_{ik} r_i \tau_{ik} + g_k(\mathbf{y_k},
\mathbf{r}) \le z_k, \forall k,
\end{equation}
\noindent where $r_i$ is the traffic demand of logical link $i$, $\tau_{ik}$ is the
unit data relay time of $i$ via relay $k$.
The first term on the left side represents the percentage of relay
capacity occupied by all the logical links using this relay as their primary
path relay.  The second term represents the \emph{protection function} for
the set of logical links using this relay as their secondary path relay.
A protection function $g_k(.)$ measures the robustness of an mmWave WPAN. Its meaning will
be further explained in Section~\ref{sec:rmrp_uncert}.

To this end, the RMRP problem can be formally stated as:
\small
\begin{equation} \label{RMRP}
\begin{aligned}
& \underset{\mathbf{x, y, z}}{\text{minimize}} & & \sum_{k} z_k \\
& \text{subject to   } & & \text{Constraints }
\eqref{eq5_x} - \eqref{scheduling} 
\\
& \text{variables} & & x_{ik}, y_{ik}, z_k \in \{0, 1\}, \forall i \in \Omega,
k = 1, \ldots, K
\end{aligned}
\end{equation}
\normalsize


\subsection{Reformulation under the $D$-norm uncertainty model}
\label{sec:rmrp_uncert}
Several uncertainty models have been proposed in literature, including
\emph{General Polyhedron}, \emph{$D$-norm}, \emph{Ellipsoid}, etc \cite{yang091}.
In this paper, we adopt the $D$-norm uncertainty model that is characterized by the following protection function:
\begin{equation} \label{DnormUncertaintyModel}
g_k(\mathbf{y_k}, \mathbf{r}) = \underset{S_k: S_k \subseteq \Omega_k, \mid S_k
\mid=\Gamma_k}{\text{max }} \sum_{i \in S_k} y_{ik} r_i \tau_{ik}.
\end{equation}
Under the $D$-norm uncertainty model, among the set of logical links $\Omega_k$ that can
relay through relay $k$, at most $\Gamma_k$ links will be blocked
simultaneously on the primary path and consequently transmit on their secondary
path via relay $k$. The maximization gives the worst case traffic loads
induced on the relay.

Two special cases are of particular interest. If $\Gamma_k = |\Omega_k|$, then
$g_k(\mathbf{y_k},\mathbf{r}) = \sum_{i \in \Omega_k} y_{ik} r_i \tau_{ik}$.
This means all logical links in $\Omega_k$ fail simultaneously. This
corresponds to the maximum robustness.  In this case, more relays may be
needed. At the other extreme, if $\Gamma_k = 0$, no logical link is blocked.
Fewer relays are in use.  However, there is little fault tolerance in the
resulting relay placement.
Denote $\rho \equiv \Gamma_k / |\Omega_k|$ the {\em robustness index}.
$\rho$ is a parameter to tradeoff between robustness and resource usage.

Under the above $D$-norm uncertainty model, \eqref{scheduling} can be rewritten as:
\small
\begin{equation} \label{scheduling_dnorm}
\sum_{i \in \Omega_k} \eta_i x_{ik} r_i \tau_{ik} + \underset{S_k: S_k
\subseteq \Omega_k, \mid S_k \mid=\Gamma_k}{\text{max }} \sum_{i \in S_k}
y_{ik} r_i \tau_{ik} \le z_k, \forall k,
\end{equation}
\normalsize

\eqref{scheduling_dnorm} is not directly tractable since it
involves an inner-optimization in the protection function. The protection
function can be reformulated an integer linear programming problem as follows
\cite{yang091}:
%
%
\begin{eqnarray} \label{scheduling_dnorm_rewrite}
\nonumber \underset{ \{0 \le s_{ik} \le 1 \}_{\forall i \in
\Omega_k}}{\text{max}} \sum_{i \in \Omega_k} y_{ik} r_i \tau_{ik} s_{ik}, \\
\text{s.t.  }  \sum_{i \in \Omega_k } s_{ik} \le \Gamma_k, \\
\nonumber s_{ik} \in \{0,1\},  \forall i \in \Omega_k
\end{eqnarray}

Consider a linear relaxation of the above problem where $s_{ik} \in [0,1]$. Due
to the linearity of the constraints, the optimal solution occurs at the
vertices of the feasibility region. Hence the optimal solution $s_{ik}^*$ must
be either $0$ or $1$, as there is no gap between the integer linear programming
and the linear programming solutions.

Taking the dual of the linear programming problem \eqref{scheduling_dnorm_rewrite},
we have:
\begin{eqnarray} \label{scheduling_dnorm_dual}
\nonumber \underset{ \{p_{ik} \ge 0 \}_{\forall i \in \Omega_k}, q_k \ge 0 }
{\text{min}} q_k \Gamma_k + \sum_{i \in \Omega_k} p_{ik}, \\ \text{s.t.  }  q_k
+ p_{ik} \ge y_{ik} r_i \tau_{ik},
\end{eqnarray}

Substituting \eqref{scheduling_dnorm_dual} into \eqref{RMRP}, we can
obtain the equivalent formulation of the RMRP problem as a \emph{mixed integer
linear programming} (MILP) problem as follows:
\small
\begin{equation} \label{RMRP_final}
\begin{aligned}
& \underset{\mathbf{x, y, z, p, q}}{\text{min}} & & \sum_{k} z_k \\
& \text{s.t.  } & & \sum_{i \in \Omega_k} \eta_i x_{ik} r_i \tau_{ik} +
q_k\Gamma_k + \sum_{i \in \Omega_k} p_{ik} \le z_k, \; \forall k \\
&&& q_k + p_{ik} \ge y_{ik} r_i \tau_{ik}, \; \forall i \in \Omega_k, \forall k
\\
&&& \text{Constraints } \eqref{eq5_x} \eqref{eq5_y}\eqref{unique_ref}
\eqref{tx_time} \eqref{rate_sr} \eqref{rate_rd}\\
%
%
%
%
& \text{variables} & & x_{ik},y_{ik},z_k \in \{0, 1\}, p_{ik} \ge 0, q_k \ge 0
\\ 
\end{aligned}
\end{equation}
\normalsize

\subsection{Hardness of RMRP}
We prove in Appendix~\ref{sec:rmrp_np} that RMRP is NP-hard.

The MILP problem in \eqref{RMRP_final} can be solved by any MILP solver. In our implementation, we
adopt the MILP solver of the IBM optimization tool -- CPLEX \cite{cplex}.

\section{Robust Maximum Utility Relay Placement (RMURP)}
\label{sec:rmurp}

In contrast to RMRP, which tries to minimize the number of relays, RMURP
aims to maximize the total utility of an mmWave WPAN given a fixed number of
relays.

Let $r_i$ be the base traffic demand on logical link $i$. We allow $r_i$ be
scaled up/down according to network constraints.  That is, the actual
data rate supported is $\alpha r_i$, where $\alpha$ is a scaling parameter.
This formulation is particularly relevant for transferring multimedia content that allows adaptive encoding.
The objective of RMURP is henceforth to maximize the total utility $U$ of the
network, given by $U = \sum_{i} \alpha r_i$.

The constraints of RMURP are similar to those of RMRP, except that the TDMA
schedulability constraint \eqref{scheduling} now becomes
\begin{equation} \label{scheduling_rmurp}
\sum_{i \in \Omega_k} \eta_i x_{ik}
\alpha r_i \tau_{ik} + g_k(\mathbf{y_k}, \alpha \mathbf{r}) \le z_k, \forall k,
\end{equation}
and the additional cardinality constraint needs to be included:
\begin{equation} \label{cardinality}
\sum_kz_k \le m,
\end{equation}
where $m$ is the maximum number of relays to be used.

To this end, the RMURP can be formalized as:
\small
\begin{equation} \label{RMURP}
\begin{aligned}
& \underset{\mathbf{x, y, z, \alpha}}{\text{maximize}} & & \sum_{i} \alpha r_i
\\
& \text{subject to} & & \text{Constraints } \eqref{eq5_x} \eqref{eq5_y}
\eqref{unique_ref} \eqref{tx_time} \eqref{rate_sr} \eqref{rate_rd}
\eqref{scheduling_rmurp} \eqref{cardinality}\\
& \text{variables} & & x_{ik}, y_{ik}, z_k \in \{0, 1\}, \alpha \ge 0, \forall
i, k 
\end{aligned}
\end{equation}
\normalsize

\subsection{Reformulation under $D$-norm Uncertainty Model}
\label{sec:rmurp_uncert}

Again we can apply the $D$-norm uncertainty model (see Section~\ref{sec:rmrp_uncert}) to the
RMURP problem of \eqref{RMURP}. This will transform \eqref{RMURP} into a
\emph{mixed integer non-linear programming problem} (MINLP) as follows:
\small
\begin{equation} \label{RMURP_final}
\begin{aligned}
& \underset{\mathbf{x, y, z, p, q, \alpha}}{\text{Maximize}} & & \sum_{i}
\alpha r_i \\
& \text{subject to  } & & \sum_{i \in \Omega_k} \eta_i x_{ik} \alpha r_i
\tau_{ik} + q_k\Gamma_k + \sum_{i \in \Omega_k} p_{ik} \le z_k, \; \forall k \\
&&& q_k + p_{ik} \ge y_{ik} \alpha r_i \tau_{ik}, \; \forall i \in \Omega_k,
\forall k \\
&&& \text{Constraints } \eqref{eq5_x} \eqref{eq5_y} \eqref{unique_ref}
\eqref{tx_time} \eqref{rate_sr} \eqref{rate_rd} \eqref{cardinality}\\
& \text{variables} & & x_{ik},y_{ik},z_k \in \{0, 1\}, p_{ik} \ge 0, q_k \ge 0,
\alpha \ge 0 \\
%
%
\end{aligned}
\end{equation}
\normalsize

\subsection{Hardness of RMURP}
%

We prove in Appendix~\ref{sec:rmurp_np} that RMURP is NP-hard. The inclusion of
variable $\alpha$ renders RMURP an MINLP. Specialized algorithms need to be designed.

Next, we propose two algorithms to solve the RMURP.  The first algorithm is
based on \emph{Bisection Search}, which is shown to be fast but does not
guarantee optimality. The second algorithm is based on the \emph{Generalized
Benders' Decomposition} (GBD) technique \cite{li2006, hua12_tmc},
which is proven to converge to optimal solutions but has higher computation
complexity.

\subsection{Bisection Search}

\emph{Bisection Search} is a heuristic method for finding roots of an equation.
It iteratively bisects an interval and then selects the subinterval where a
root must reside for the next iteration, until some termination condition is
met.  It is guaranteed to converge to a root of $f(\cdot)$ if and only if: $f$
is a continuous function on the interval $[A,B]$, and $f(A)$ and $f(B)$ have
opposite signs.

In \eqref{RMURP_final}, if $\alpha$ is given, RMURP becomes a MILP, which can be
solved by CPLEX. The key is thus to determine the value of $\alpha$. When
$\alpha$ is large, RMURP is infeasible. When $\alpha = 0$, RMURP is always
feasible. More generally, if $\alpha_1 > \alpha_2$, RMURP is infeasible when
$\alpha = \alpha_1$, then, RMURP is feasible when $\alpha = \alpha_2$.  Treating
feasibility and infeasibility as opposite signs, we apply the \emph{Bisection Search}
principle to decide the range of $\alpha$ iteratively until it is smaller
than a threshold $2\times TOL$. Starting from an initial interval $[A, B]$,
where $\alpha = A$ renders RMURP feasible and $\alpha = B$ renders RMURP
infeasible, we substitute A or B with $\frac{A+B}{2}$ depending on the
feasibility of RMURP under $\alpha = \frac{A+B}{2}$. The ``monotonicity" in the
feasibility of RMURP with respective to $\alpha$ makes the \emph{Bisection Search}
converge fast but the optimality of the final results depends on $TOL$.

The \emph{Bisection Search}  based algorithm is summarized in Algorithm~\ref{BisectionAlgo}.
\begin{algorithm}
\footnotesize
\SetAlgoLined
\DontPrintSemicolon
\SetKwData{Left}{left}\SetKwData{This}{this}\SetKwData{Up}{up}
\SetKwFunction{Union}{Union}\SetKwFunction{FindCompress}{FindCompress}
\SetKwInOut{Input}{Input}\SetKwInOut{Output}{Output}
\Input{Base logical data rate $r_i$ for each feasible logical link $i$;
error tolerance $TOL$ (which serves as the iteration termination condition);
and the up-to-date known range for $\alpha$: [$A$, $B$]}
\Output{Maximum network utility $U_T$ and relay selection variables for
every feasible links $\mathbf{x, y, z}$ }
\Begin{
    Set $n=1$; \;
    \While{$n \le maxN$}{
	 $C \longleftarrow \frac{(A+B)}{2}$; \;
	 Solve the MILP problem $RMURP(C)$. \;
	 \uIf{$RMURP(C)$ is feasible}
	    {$A \longleftarrow C$; \;
	     Obtain the solutions of $RMURP(C)$: $\mathbf{x}^{(n)}$, $\mathbf{y}^{(n)}$, $\mathbf{z}^{(n)}$.}
	 \Else{$B \longleftarrow C$; \;}
	 $n \longleftarrow n+1$; \;
	 \If{$\frac{(B-A)}{2} \le TOL$}
	    {The best $\alpha$ found, $\alpha_{best} \longleftarrow A$; \;
	     Return $U_T$, $\mathbf{x}^{(n)}$, $\mathbf{y}^{(n)}$, $\mathbf{z}^{(n)}$. \; }
    }
}
\caption{Bisection Search\label{BisectionAlgo}}
\normalsize
\end{algorithm}


\subsection{Generalized Benders' Decomposition (GBD)}

\emph{GBD} is an iterative method for solving MINLP problems.
The principle of the \emph{GBD} algorithm is to decompose the original MINLP problem into a \emph{primal problem} and a
\emph{master problem}, and then solve them iteratively.  The primal problem
corresponds to the original problem with fixed binary variables. Solving the
primal problem provides a lower bound, and Lagrange multipliers
corresponding to the constraints.  The master problem is derived through
nonlinear duality theory using the Lagrange multipliers obtained from the
primal problem.  Solving the master problem provides a upper bound,
and binary variables that can be used for the primal problem in the next iteration.
It is proven to converge to the optimum~\cite{li2006}.

\paragraph*{Primal Problem} Let $\Lambda := (\mathbf{x}, \mathbf{y}, \mathbf{z})$ represent the set of
binary variables, $\hat{\Lambda}:= (\mathbf{\hat{x}}, \mathbf{\hat{y}}, \mathbf{\hat{z}})$ indicates the
binary variables with specific values in $\{0, 1\}$.  The primal problem
$\mathcal{P}(\hat{\Lambda})$ of RMURP problem \eqref{RMURP_final} is obtained
by fixing all the binary variables to $\hat{\Lambda}$ as follows:
\small
\begin{equation} \label{GBD_primal2}
%
%
\begin{aligned}
f(\hat{\Lambda}) = & \underset{\mathbf{p}, \mathbf{q}, \alpha}{\text{maximize}}
& & \sum_{i} \alpha r_i \\
& \text{subject to} & & \sum_{i \in \Omega_k} \eta_i \hat{x}_{ik} \alpha r_i
\tau_{ik} + q_k\Gamma_k + \sum_{i \in \Omega_k} p_{ik} \le \hat{z}_k, \;
\forall k \\
&&& q_k + p_{ik} \ge \hat{y}_{ik} \alpha r_i \tau_{ik}, \; \forall i \in
\Omega_k, \forall k \\
%
& \text{variables} & & \mathbf{p} \succeq 0, \mathbf{q} \succeq 0, \alpha \ge 0
\\
\end{aligned}
%
%
\end{equation}
\normalsize

Problem \eqref{GBD_primal2} is a linear programming problem, which can be
solved by any linear programming solver. Since the optimal solution of
$\mathcal{P}(\hat{\Lambda})$ is also a feasible solution to
\eqref{RMURP_final}, the optimal value $f(\hat{\Lambda})$ provides a
lower bound to the RMURP.  It is also clear that, not all the choices
of given binary variables can lead to a feasible primal problem.  We need to
treat it differently depending on whether the primal problem is feasible or
not:

$\bullet$ Feasible Primal:

If the primal problem is feasible, let
\small
\begin{equation} \label{}
\begin{aligned}
& T_k(\hat{\Lambda}, \mathbf{p}, \mathbf{q}, \alpha) = \hat{z}_k - (\sum_{i \in \Omega_k} \eta_i \hat{x}_{ik} \alpha r_i \tau_{ik} + q_k\Gamma_k + \sum_{i \in \Omega_k} p_{ik}), \forall k, \\
& g_{ik}(\hat{\Lambda}, \mathbf{p}, \mathbf{q}, \alpha) = q_k + p_{ik} -\hat{y}_{ik} \alpha r_i \tau_{ik}, \forall i \in \Omega_k, \forall k.\\
\end{aligned}
\end{equation}
\normalsize
Then, we can compute the \emph{partial Lagrangian} function for the primal problem as follows:
\small
\begin{equation} \label{}
L(\hat{\Lambda}, \mathbf{p}, \mathbf{q}, \alpha, \mathbf{\lambda}, \mathbf{\nu} ) = \sum_{i} \alpha r_i + \sum_{k} \lambda_k T_k + \sum_{k} \sum_{i} \nu_{ik} g_{ik},
\end{equation}
\normalsize
where $\lambda_k, \nu_{ik} \ge 0, \forall i \in \Omega_k, \forall k$ are the \emph{Lagrange multipliers}.

Thus, the \emph{Lagrange dual} problem of $\mathcal{P}(\hat{\Lambda})$ can be
stated as:
\begin{equation} \label{primal_feasible_dual}
\underset{\mathbf{\lambda}, \mathbf{\nu} }{\text{min }} \underset{p, q,
\alpha}{\text{ max }} L(\hat{\Lambda}, \mathbf{p}, \mathbf{q}, \alpha,
\mathbf{\lambda}, \mathbf{\nu}).
\end{equation}
Since the problem is convex and has linearity constraints,
the duality gap is 0. Thus, solving the Lagrange dual problem would give the
optimal solution for $\mathcal{P}(\hat{\Lambda})$.

$\bullet$ Infeasible Primal:

If the primal problem is infeasible, we first define a set $\Delta$ as:
\small
\begin{equation}
\Delta = \{ \hat{\Lambda} | T_k \ge 0, g_{ik} \ge 0, \forall i \in \Omega_k,
\forall k, \text{for some } \mathbf{p}, \mathbf{q}, \alpha \},
\end{equation}
\normalsize
and consider the following feasibility-checking problem
$\mathcal{F}(\hat{\Lambda})$:
\small
\begin{equation} \label{primal_check}
%
%
\begin{cases}
\begin{aligned}
& \underset{\mathbf{p}, \mathbf{q}, \alpha}{\text{minimize}} & & \delta \\
& \text{subject to} & & \sum_{i \in \Omega_k} \eta_i \hat{x}_{ik} \alpha r_i
\tau_{ik} + q_k\Gamma_k + \sum_{i \in \Omega_k} p_{ik} - \hat{z}_k \le \delta,
\; \forall k \\
&&& \hat{y}_{ik} \alpha r_i \tau_{ik} - q_k - p_{ik} \le \delta , \; \forall i
\in \Omega_k, \forall k \\
%
%
& \text{variables} & & \mathbf{p} \succeq 0, \mathbf{q} \succeq 0, \alpha \ge
0, \delta \ge 0 \\
\end{aligned}
\end{cases}
\end{equation}
\normalsize

It is straightforward to see that, for any given $\hat{\Lambda}$,
$\mathcal{P}(\hat{\Lambda})$ is infeasible if and only if
$\mathcal{F}(\hat{\Lambda})$ has a positive optimal value $\delta^* > 0$.

The \emph{Lagrangian} function for $\mathcal{F}(\hat{\Lambda})$ can be
presented as:
\small
\begin{equation} \label{}
\begin{aligned}
G(\hat{\Lambda}, \mathbf{p}, \mathbf{q}, \alpha, \mathbf{\mu}, \mathbf{\sigma})
= & \sum_{k} \mu_k (\sum_{i \in \Omega_k} \eta_i \hat{x}_{ik} \alpha r_i
\tau_{ik} + q_k\Gamma_k + \sum_{i \in \Omega_k} p_{ik} \\
& - \hat{z}_k) + \sum_{k} \sum_{i \in \Omega_k} \sigma_{ik} (\hat{y}_{ik}
\alpha r_i \tau_{ik} - q_k - p_{ik}), \\
& \forall (\mu_k, \sigma_{ik}) \in \mho \\
\end{aligned}
\end{equation}
\normalsize
where $\mu_k, \sigma_{ik}$ are Lagrange multipliers and $\mho = \{ (\mu_k,
\sigma_{ik}) | \mu_k, \sigma_{ik} \ge 0, \sum_{k} (\mu_k + \sum_{i \in
\Omega_k} \sigma_{ik})= 1, \forall i \in \Omega_k, \forall k \}$.

The Lagrangian dual of $\mathcal{F}(\hat{\Lambda})$ becomes:
\begin{equation} \label{}
\underset{ \mu, \sigma }{\text{max }} \underset{\mathbf{p}, \mathbf{q},
\alpha}{\text{ min }} G(\hat{\Lambda}, \mathbf{p}, \mathbf{q}, \alpha,
\mathbf{\mu}, \mathbf{\sigma}).
\end{equation}

Therefore, for any $\hat{\Lambda} \in \Delta$, it can be characterized by the
inequality constraint:
\begin{equation} \label{infea_const}
0 \ge \underset{p, q}{\text{min }}  G(\hat{\Lambda}, \mathbf{p}, \mathbf{q},
\alpha, \mathbf{\mu}, \mathbf{\sigma}).
\end{equation}

\paragraph*{Master Problem}

The original problem in \eqref{RMURP_final} can be written as:  \\
\small \begin{equation} \label{GBD_master2}
\begin{aligned} &
\underset{\Lambda}{\text{max }} \sum_{i} \alpha r_i && = \underset{\Lambda \in
\Delta}{\text{max }}  f(\Lambda) \\
&&& = \underset{\Lambda \in
\Delta}{\text{max }} \left[ \underset{\lambda, \nu}{\text{min }}
\underset{\mathbf{p}, \mathbf{q}, \alpha}{\text{max }} L(\Lambda, \mathbf{p},
\mathbf{q}, \alpha, \mathbf{\lambda}, \mathbf{\nu} ) \right] \\
&&& =
\underset{}{\text{max }} \beta \\
&&& \hspace{6mm} \text{s.t. } \beta \le
\underset{\mathbf{p}, \mathbf{q}, \alpha}{\text{max }} L(\Lambda, \mathbf{p},
\mathbf{q}, \alpha, \mathbf{\lambda}, \mathbf{\nu} ), \forall \mathbf{\lambda}, \mathbf{\nu}
\succeq 0 \\
&&& \hspace{12mm} \Lambda \in \{0, 1\} \cap \Delta, \\
\end{aligned}
\end{equation}
\normalsize
where the second equality is due to \eqref{primal_feasible_dual} because of the zero
duality gap.  Incorporating \eqref{infea_const} into \eqref{GBD_master2}, we
finally obtain the master problem $\mathcal{M}(\mathbf{p}, \mathbf{q}, \alpha,
\lambda, \nu, \mu, \sigma)$ as:
\small
\begin{equation} \label{eq23}
\mathcal{M}(.)
\begin{cases}
\begin{aligned}
& \underset{\Lambda}{\text{max }} & & \beta \\
& \text{s.t. } & & \beta \le \underset{\mathbf{p}, \mathbf{q},
\alpha}{\text{max }}  L(\Lambda, \mathbf{p}, \mathbf{q}, \alpha,
\mathbf{\lambda}, \mathbf{\nu}), \forall \mathbf{\lambda}, \mathbf{\nu} \succeq 0 \\
& & & 0 \ge \underset{\mathbf{p}, \mathbf{q}, \alpha}{\text{min }}  G(\Lambda,
\mathbf{p}, \mathbf{q}, \alpha, \mathbf{\mu}, \mathbf{\sigma}), \forall (\mathbf{\mu}, \mathbf{\sigma}) \in \mho \\
& & &  \text{Constraints } \eqref{eq5_x} \eqref{eq5_y} \eqref{unique_ref}
\eqref{tx_time} \eqref{rate_sr} \eqref{rate_rd} \eqref{cardinality} \\
& & &  \Lambda \in \{0, 1\}, \beta \ge 0 \\
\end{aligned}
\end{cases}
\end{equation}
\normalsize

Note that, the master problem has two inner optimization problems as its
constraints, which need to be considered for all $\mathbf{\lambda}$,
$\mathbf{\nu}$ and $\mathbf{\mu}$, $\mathbf{\sigma}$.  This implies that the
master problem has a very large number of constraints.  In order to obtain a
solvable mixed-integer linear programming problem, we employ the following
relaxation for the master problem at iteration $n$ as suggested
by~\cite{li2006}:
\small
\begin{equation} \label{eq21}
\begin{aligned}
& \beta \le L(\Lambda^n, \mathbf{p}^n, \mathbf{q}^n, \alpha^n,
\mathbf{\lambda}^n, \mathbf{\nu}^n) + \bigtriangledown_{\Lambda} L(.)(\Lambda -
\Lambda^n), \forall n \in \mathcal{P}^k \\
& 0     \ge G(\Lambda^n, \mathbf{p}^n, \mathbf{q}^n, \alpha^n, \mathbf{\mu}^n,
\mathbf{\sigma}^n) + \bigtriangledown_{\Lambda} G(.)(\Lambda - \Lambda^n),
\forall n \in \mathcal{F}^k, \\
\end{aligned}
\end{equation}
\normalsize
where $\mathcal{P}^k$ and $\mathcal{F}^k$ are the sets of feasible and
infeasible primal problems solved up to iteration $k$, respectively.

The relaxed problem provides the upper bound of the original master problem and
also provide the value of binary variables for the primal problem in the next
iteration.

The \emph{GBD} algorithm is summarized in Algorithm~\ref{GBDalgo}.

\begin{algorithm}
\footnotesize
\SetAlgoLined
\DontPrintSemicolon
\SetKwData{Left}{left}\SetKwData{This}{this}\SetKwData{Up}{up}
\SetKwFunction{Union}{Union}\SetKwFunction{FindCompress}{FindCompress}
\SetKwInOut{Input}{Input}\SetKwInOut{Output}{Output}
\Input{Base logical data rate $r_i$ for each feasible logical link $i$}
\Output{Maximum network utility, adapt factor $\alpha$ and relay selection
variables for every feasible logical link $\Lambda =(\mathbf{x, y, z})$ and
$\mathbf{p}, \mathbf{q}$ }
\Begin{
    set $n=1$ and choose $\Lambda \in \{0, 1\}, $ \;
    $LB^0 \longleftarrow -\infty, UB^0 \longleftarrow \infty, \mathcal{P}^0
\longleftarrow \emptyset, \mathcal{F}^0 \longleftarrow \emptyset. $\;
    \While{$LB^{n-1} \le UB^{n-1} $}{
	\uIf{the primal problem is feasible} {
	Solve the primal problem $\mathcal{P}(\Lambda^n)$ to obtain optimal
solution $\mathbf{p}^n, \mathbf{q}^n, \alpha^n $ \;
	and Lagrangian multipliers $\mathbf{\lambda}^n, \mathbf{\nu}^n $; \;
	$\mathcal{P}^n \longleftarrow \mathcal{P}^{n-1} \cup \{n\}$,
$\mathcal{F}^n \longleftarrow \mathcal{F}^{n-1}$; \;
	$LB^n \longleftarrow max(LB^{n-1}, f(\Lambda^n))$; \;
	    \If{$LB^n == f(\Lambda^n)$}{
		$(\Lambda^*, \mathbf{\tilde{p}}^*, \mathbf{\tilde{q}}^*,
\mathbf{\tilde{\alpha}}^*) \longleftarrow (\Lambda^n, \mathbf{\tilde{p}}^n,
\mathbf{\tilde{q}}^n, \mathbf{\tilde{\alpha}}^n) $; \;}
	}
	\ElseIf{the primal problem is infeasible} {
	Solve the feasibility-check problem $\mathcal{F}(\hat{\Lambda})$ to
obtain the optimal solution $\mathbf{p}^n, \mathbf{q}^n, \alpha^n $
	and Lagrangian multipliers $\mathbf{\mu}^n, \mathbf{\sigma}^n $; \;
	$\mathcal{P}^n \longleftarrow \mathcal{P}^n $, $\mathcal{F}^n
\longleftarrow \mathcal{F}^{n-1} \cup \{n\}$; \;
	}
	Solve the master problem $\mathcal{M}(\mathbf{p}^n, \mathbf{q}^n,
\alpha^n, \lambda^n, \nu^n, \mu^n, \sigma^n)$ \;
	and obtain the optimal solution $\Lambda^{n+1}$ and $\beta^n$; \;
	$UB^n \longleftarrow \beta^n, n \longleftarrow n+1; $ \;
    }
    return $\Lambda^*, \mathbf{p}^*, \mathbf{q}^*, \alpha^* $.
}
\caption{Generalized Benders' Decomposition \label{GBDalgo}}
\normalsize
\end{algorithm}

It has been be proved in \cite{li2006} that, the solutions to discrete variables
$\Lambda^{1}, \ldots, \Lambda^{k}$ do not repeat.  Therefore, due to the finiteness of the discrete
variable set, \emph{GBD} algorithm converges within a finite number of iterations.
When it converges, the lower bound is equal to the upper bound. Thus,
optimality is achieved.

\section{Performance Evaluation}
\label{sec:performance}

\begin{table}[tbh!]
\begin{center}
\caption{PHY Parameters}
\label{tab:parameters}
\footnotesize
\begin{tabular}{l*{6}{l}r}
\hline
PHY parameters       & Values  \\
\hline
Channel              & AWGN with gain 1  \\
Path Loss            & free space, exponent 2 \\
Transmission power       & 20mW (13dBm)  \\
Noise floor          & -100dBm  \\
\hline
\end{tabular}
\end{center}
\end{table}
\normalsize

In this section, we evaluate the performance of the relay placement
solutions using simulations.  In the simulations, an mmWave home network is
deployed in a 10m$\times$10m room, where $N$ mmWave devices and $O$ obstacles
are uniformly placed.  The relays can be placed at any grid point in a grid
separated by distance $d_0$.  The transmission radii of all mmWave devices
and relays are set to 6 meters.  The (base) traffic demand $r_i$ of each
logical link $i$ is chosen as $\frac{1}{3}$ of the AWGN Shannon channel
capacity of the slowest LOS path. The PHY parameters are given in
Table~\ref{tab:parameters}. In all experiments, $d_0 = 2m$, $O=10$.

\subsection{RMRP Performance}

In this section, we examine the performance of RMRP under different
configurations by varying the number of mmWave logical links ($N$), the number of
moving subjects ($M$) and the robustness index ($\rho$).

\begin{figure*}[tb]
  \centering
\begin{tabular}{p{2.2in}p{2.2in}p{2.2in}}
\includegraphics[width=2.2in]{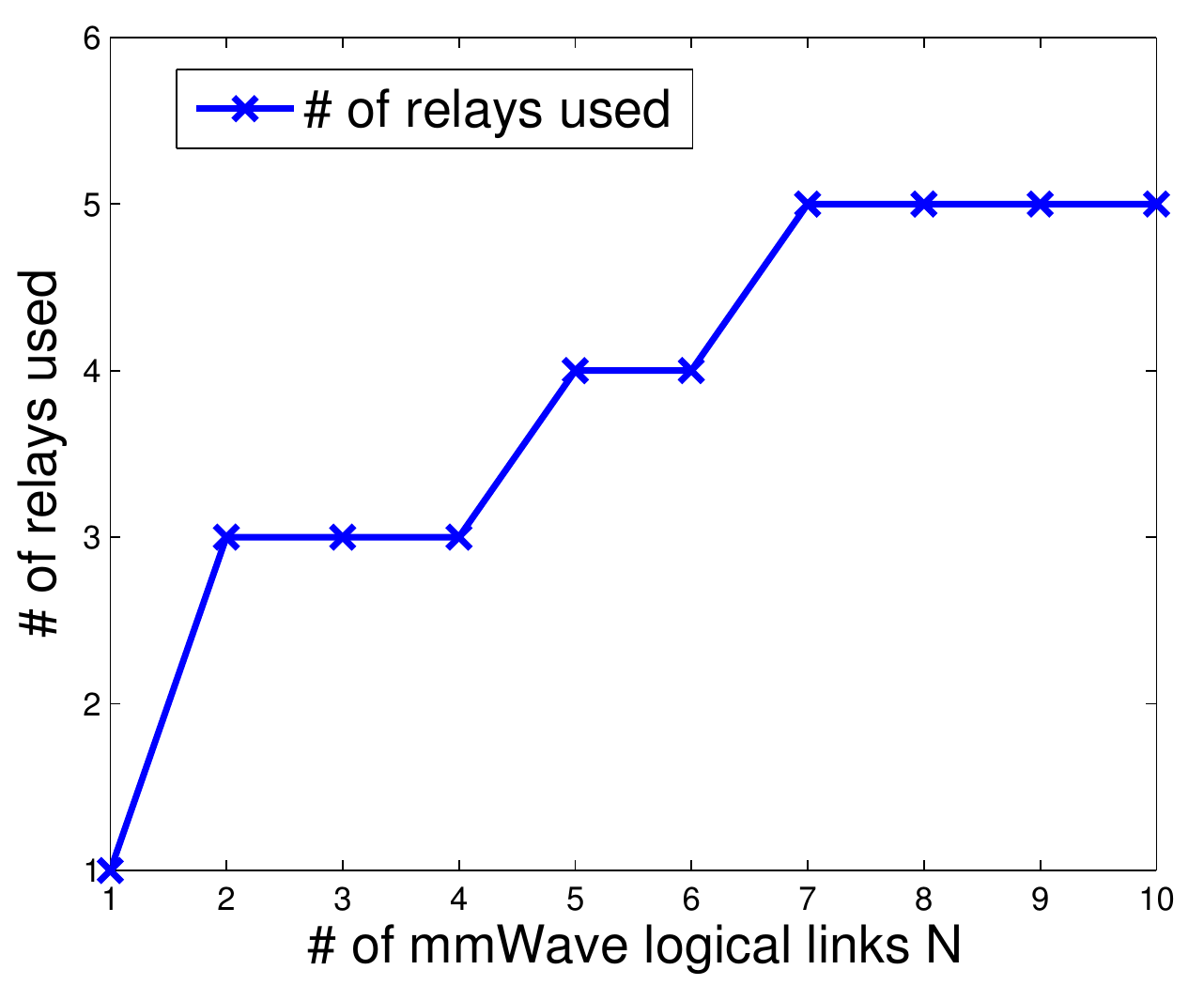}&
\includegraphics[width=2.2in]{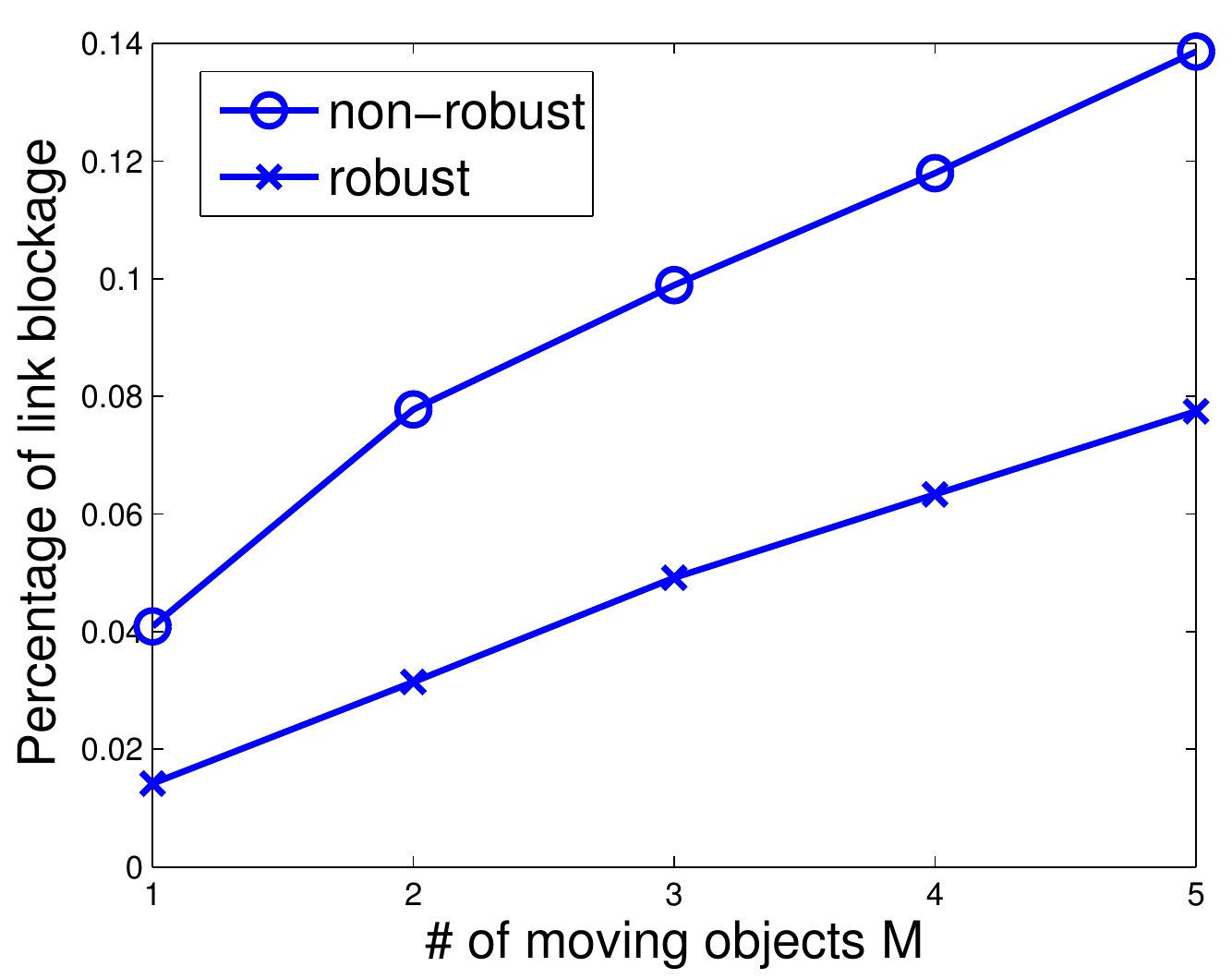}&
\includegraphics[width=2.3in]{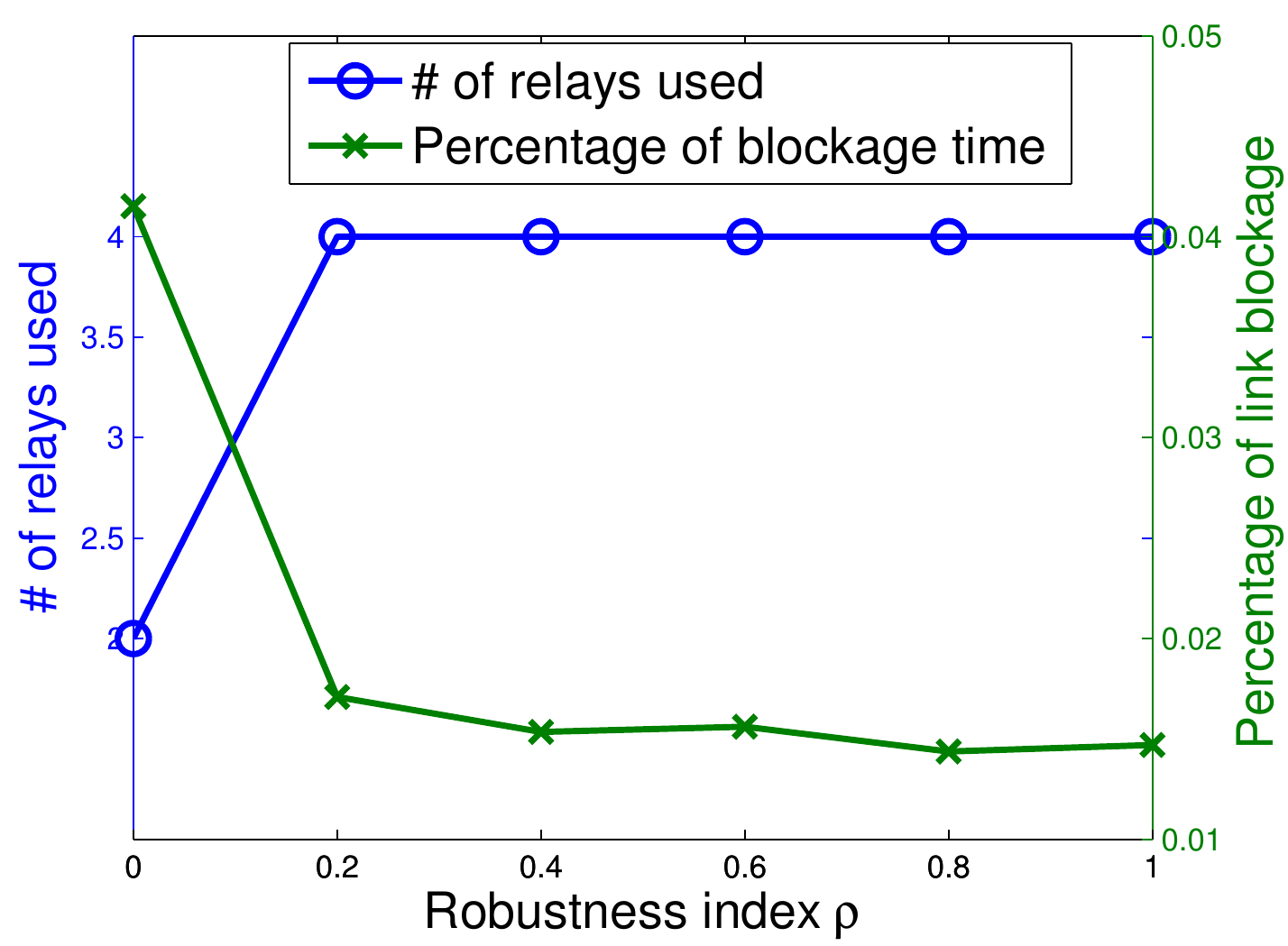}\\
(a) Effect of the number of logical links with $\rho = 1$
& (b) Effect of the number of moving subjects with $\rho = 1$, $N = 5$
& (c) Effect of the robustness index with $M = 1, N = 5$
\end{tabular}
\caption{Performance of RMRP in an mmWave home network deployed in a
10m$\times$10m room.}
\label{fig:RMRP_performance}
\end{figure*}

Fig.~\ref{fig:RMRP_performance}(a) shows the number of relays when
relay candidate locations and obstacles are fixed, while the number of
mmWave logical links varies. In this set of experiments, all logical links are
feasible. More relays are needed as the number of logical links increases.
However, the relationship is not always linear due to the absence of LOS paths
between TX/RX pairs and the multiplexing of relays.

Fig.~\ref{fig:RMRP_performance}(b) show the percentage of link blockage per
link when human subjects move randomly in the room. The mobility setup is
similar to that in Section~\ref{sec:robust}. Clearly, As the number of human
subjects increases, the percentage of link blockage increases as well. However,
the robust scheme leads to 50\% less blockage.

Next, we evaluate the impact of robustness index $\rho$. In this setup, 1 human
subject moves randomly and there are 5 logical links.
Figure~\ref{fig:RMRP_performance}(c) shows the number of relays used and
the percentage of link blockage. As expected, as $\rho$ increases, more
relays are used and the link blockage reduces.
\subsection{RMURP Performance}

We now evaluate the performance of the \emph{Bisection Search} and \emph{GBD} algorithm on
RMURP. The error tolerance of \emph{Bisection Search}  is set to $TOL = 1.0$.

\begin{figure*}[tb]
  \centering
\begin{tabular}{p{2.3in}p{2.3in}p{2.3in}}
\includegraphics[width=2.1in]{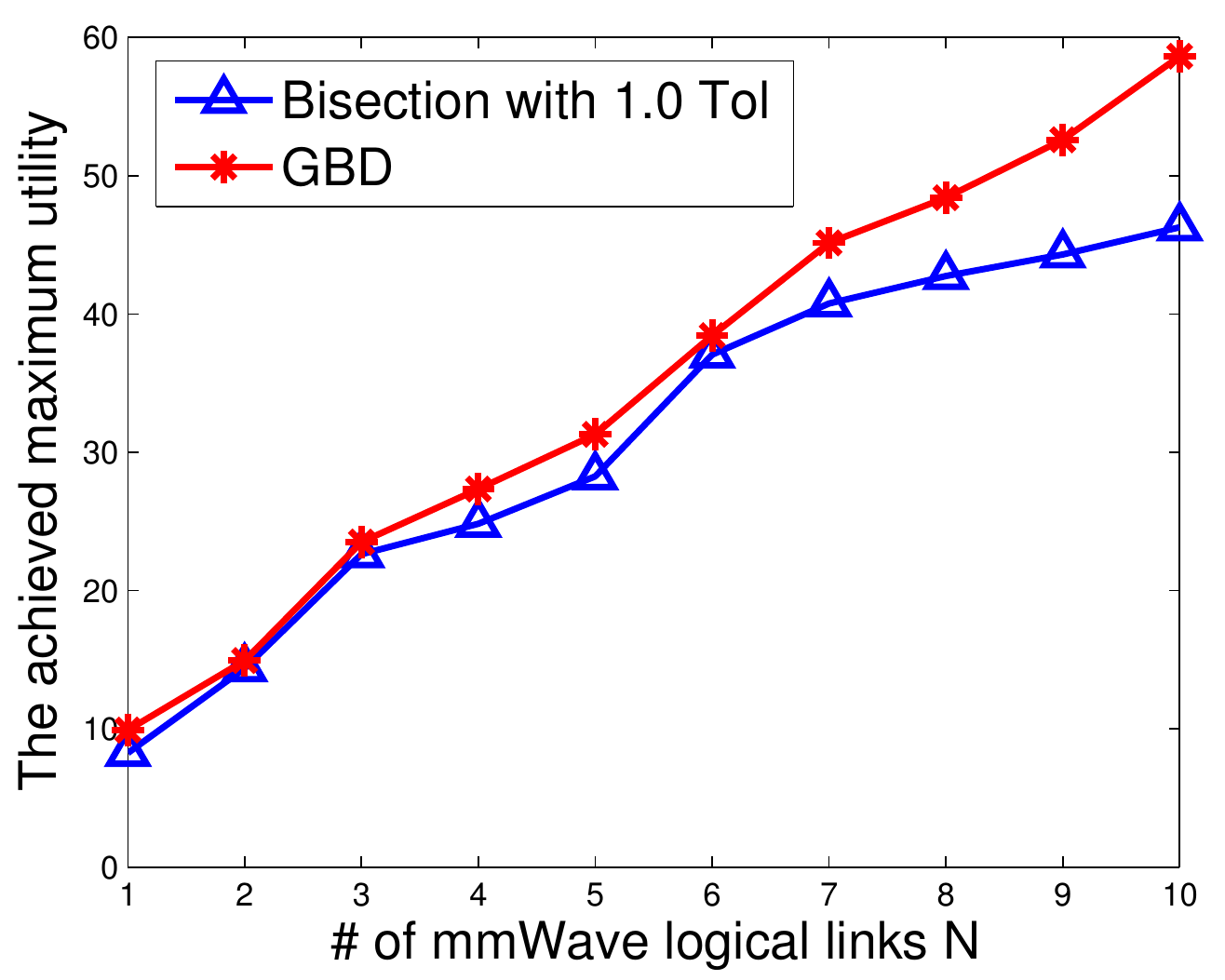} &
\includegraphics[width=2.1in]{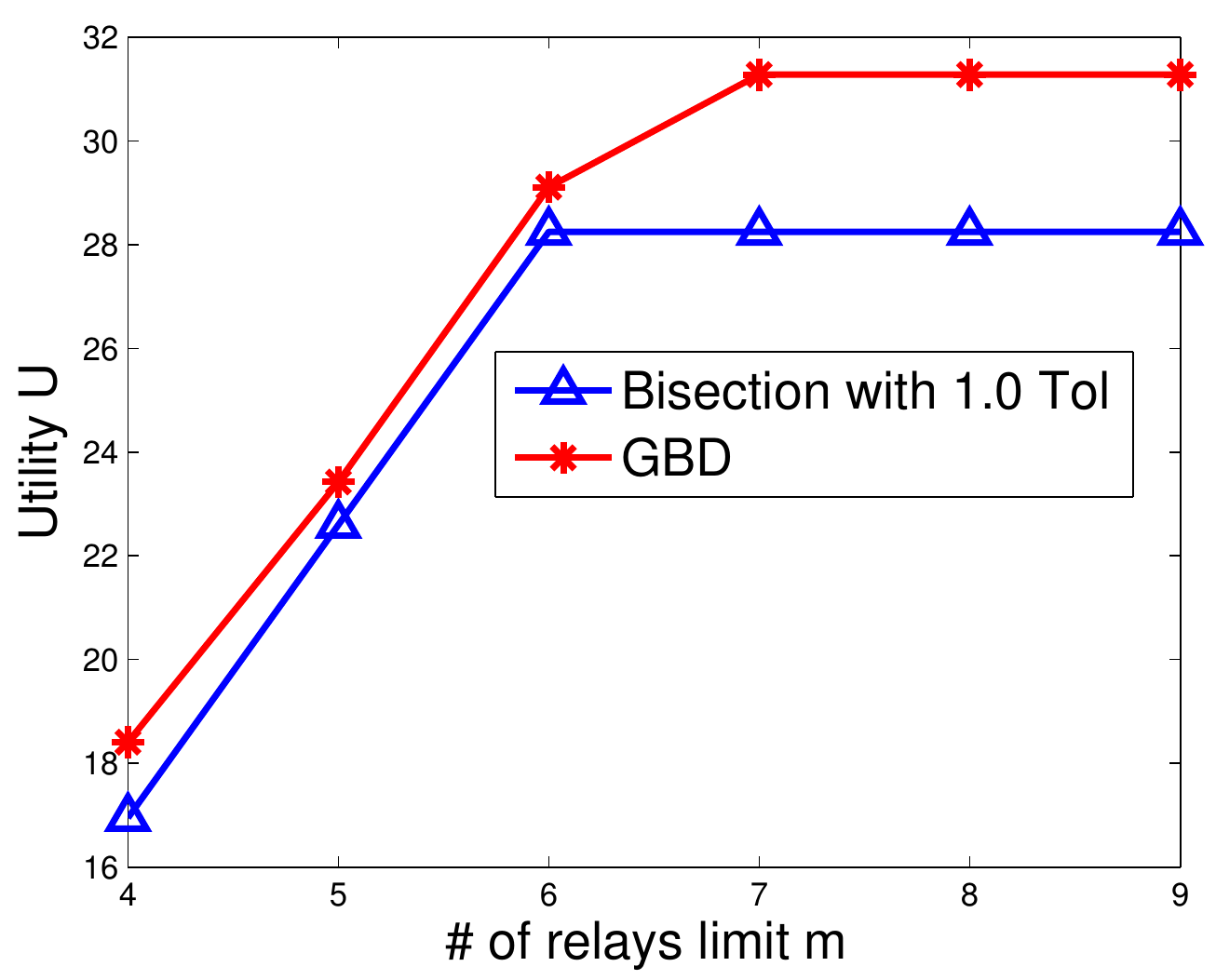} &
\includegraphics[width=2.1in]{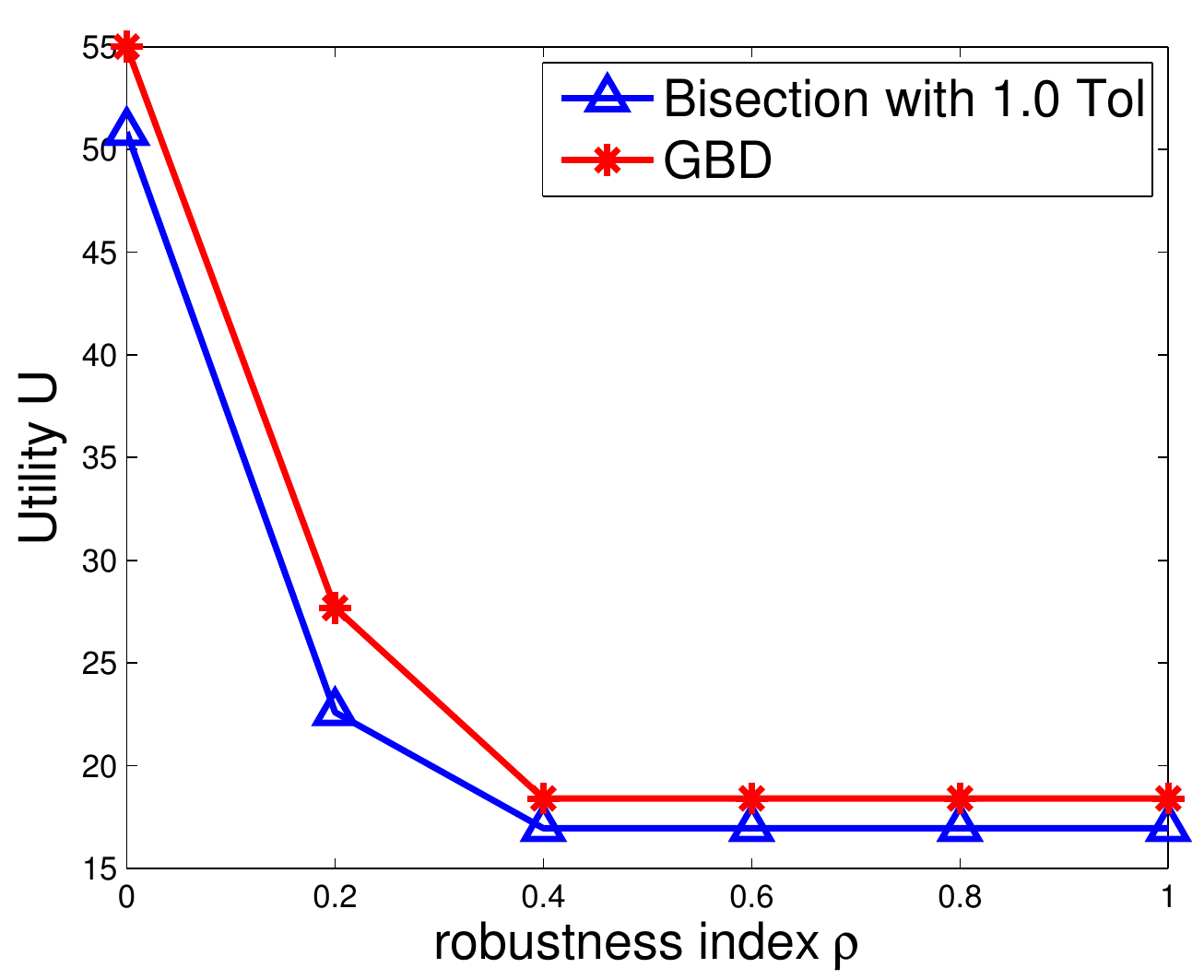} \\
(a) Effect of the number of logical links with $m=7$ and $\rho=1$
& (b) Effect of the maximum number of relays with $N=5$ and $\rho=1$
& (c) Effect of the robustness index with $m=7$ and $N = 5$ \\
%
\end{tabular}
\caption{Performance of two RMURP solutions in an mmWave home network deployed in a 10m$\times$10m room,
\emph{Bisection Search} algorithm with 1.0 tolerance and \emph{GBD} algorithm.}
\label{fig:RMURP_performance}
\end{figure*}

Fig.~\ref{fig:RMURP_performance} shows the utility achieved using both methods
by varying the number of logical links (Fig.~\ref{fig:RMURP_performance}(a)),
the total number of relays (Fig.~\ref{fig:RMURP_performance}(b)) and the
robustness index (Fig.~\ref{fig:RMURP_performance}(c)).  In all cases, \emph{GBD}
achieves higher utility compared to \emph{Bisection Search}. Reducing the
threshold $TOL$ improves the performance of \emph{Bisection Search} but comes at a
higher computation cost.
\begin{figure*}[tb]
  \centering
\begin{tabular}{p{2.3in}p{2.3in}p{2.3in}}
\includegraphics[width=2.1in]{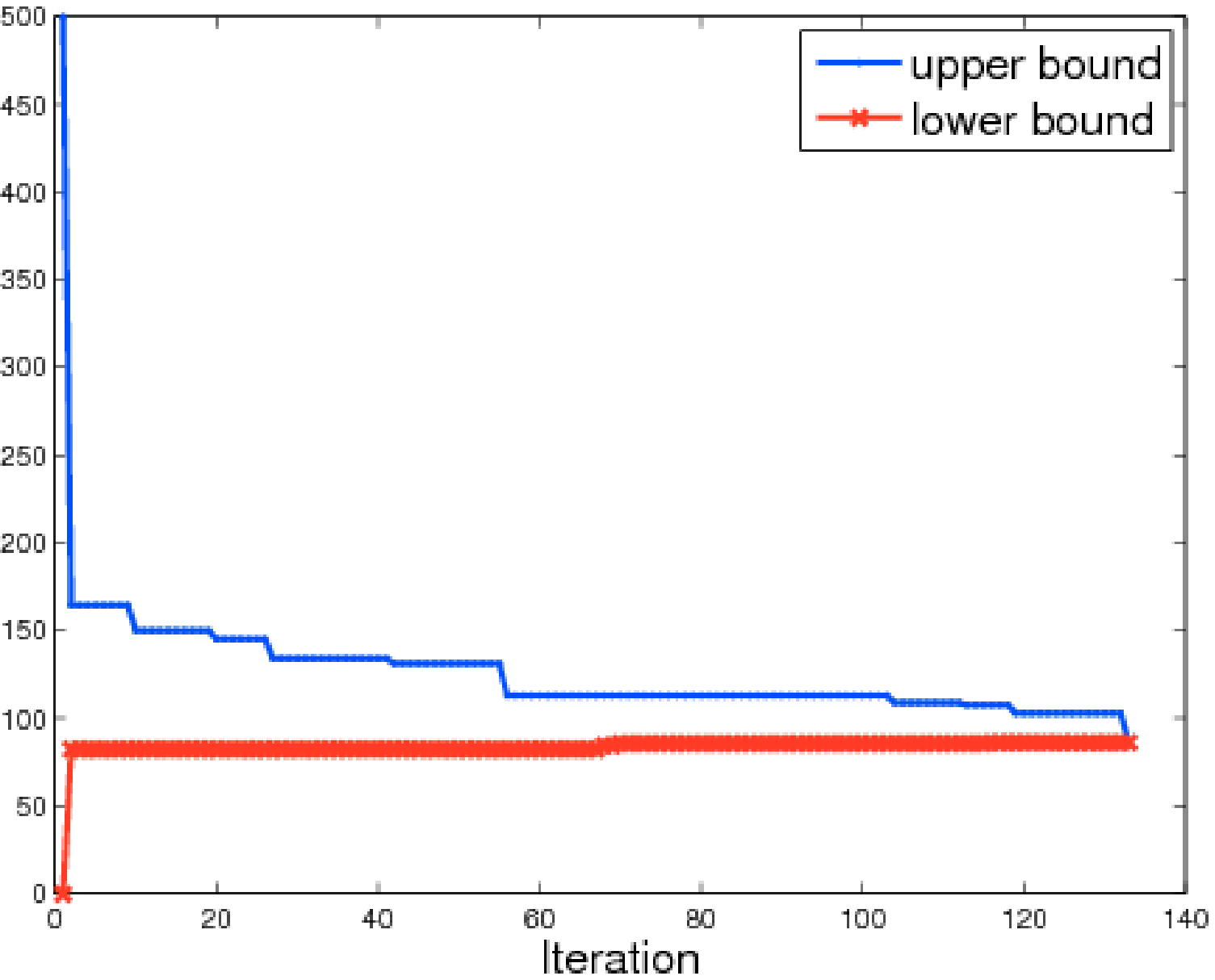} &
\includegraphics[width=2.1in]{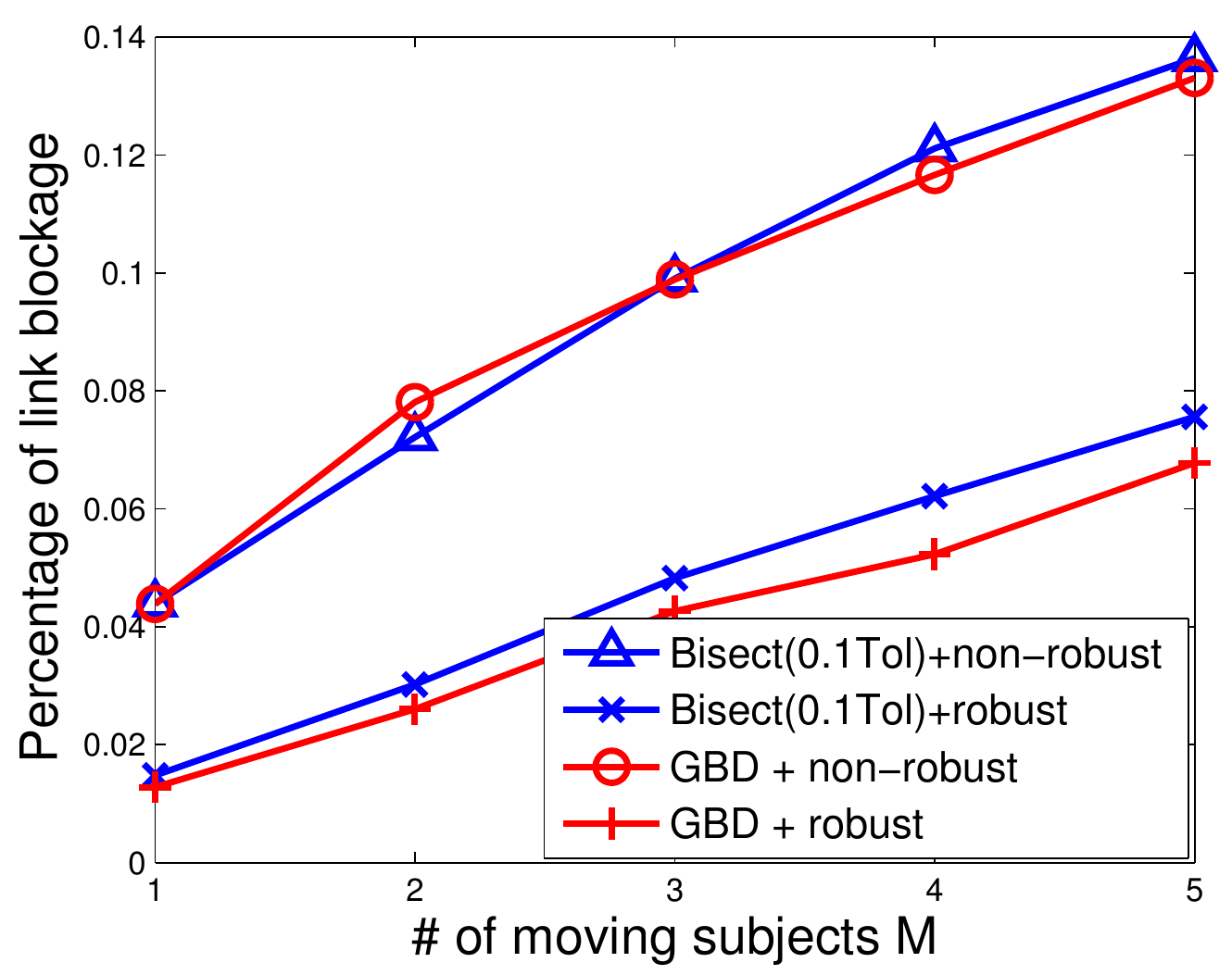} &
\includegraphics[width=2.1in]{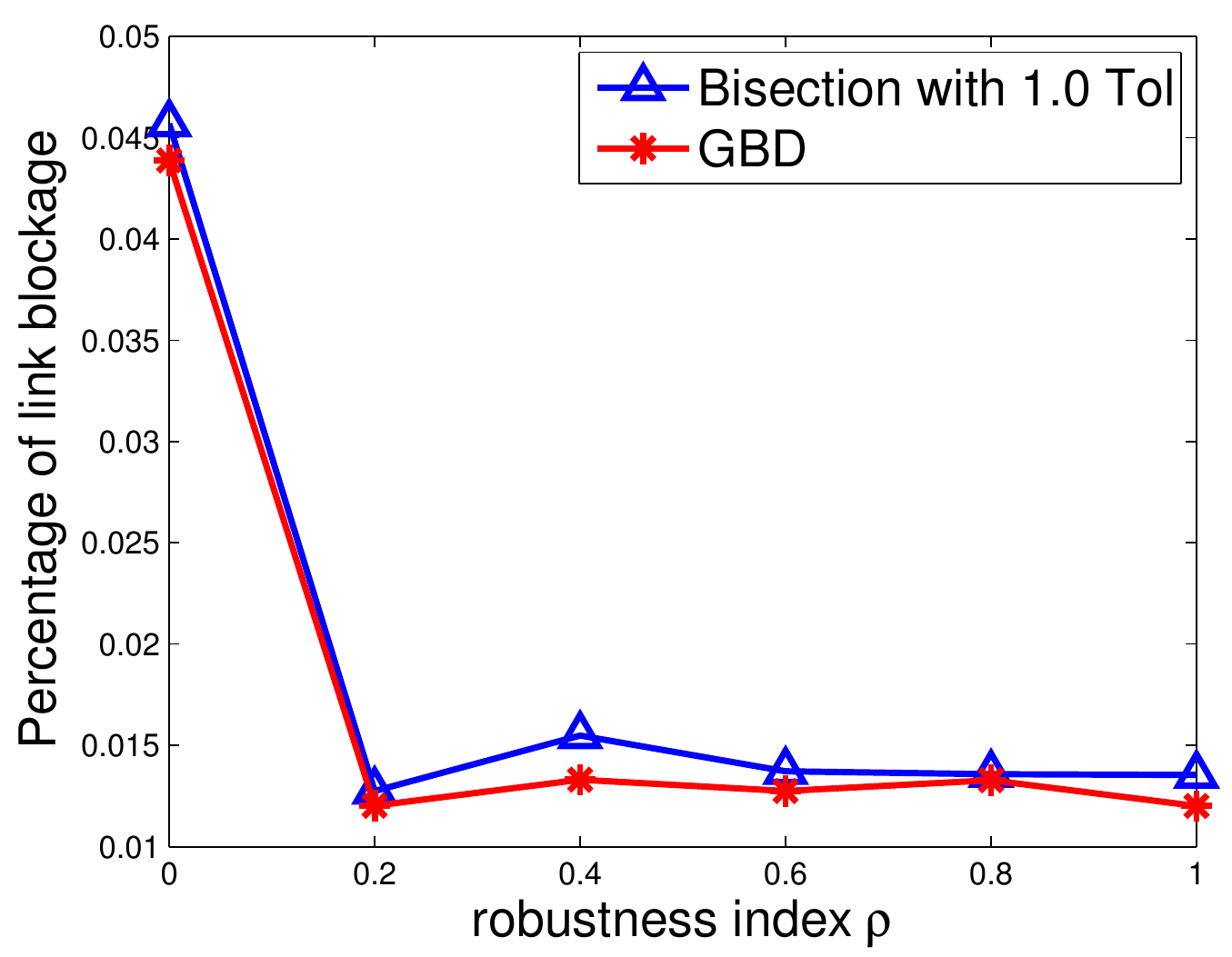} \\
(a) \emph{GBD} convergence with $N = 5$, $m = 7$ and $\rho = 1$
& (b) Effect of the number of moving subjects with $N = 5$, $m = 7$, and $\rho = 1$
& (c) Effect of the robustness index with $N=5$, $m = 7$, and $M = 1$  \\
\end{tabular}
\caption{Performance of RMURP in an mmWave home network deployed in a 10m$\times$10m room.
}
\label{fig:RMURP_performance2}
\end{figure*}

Fig.~\ref{fig:RMURP_performance2}(a) demonstrates the convergence of the \emph{GBD}
algorithm. As shown in Fig.~\ref{fig:RMURP_performance2}(a), over time, the
upper bound (solutions to the master problem) is non-increasing; and the lower
bound (solutions to the primary problem) is non-decreasing. The algorithm
converges to the optimal solution after 50 iterations when the upper bound
equals to the lower bound.

Fig.~\ref{fig:RMURP_performance2}(b)(c) shows the percentage of link blockage per link
under RMURP when human subjects move randomly in the room
by varying the number of moving subjects and the robustness index, respectively.
In both cases, \emph{GBD} achieves lower percentage of blockage. This implies that the
relay selection in \emph{GBD} has more spatial diversity.

\section{Conclusion}
\label{sec:conclusion}

In this paper, we formulated two robust relay placement problems in mmWave WPANs,
namely, robust minimum relay placement problem (RMRP) and robust
maximum utility relay placement (RMURP) for better connectivity and
robustness against link blockage. Under the D-norm uncertainty model, RMRP
and RMURP were casted as MILP and MINLP problems. Efficient algorithms were
devised and evaluated using simulations.

For future work, we will incorporate more complex interference and antenna
models in the robustness formulation.  Another interesting agenda is to explore
the use of passive relays in mmWave WPANs.

\section*{Acknowledgment}
This work is supported by the National Science Foundation under Grant number
CNS-0832084, CNS-1117560, National Natural Science Foundation of China (No.
61001096), and Hong Kong RGC General Research Fund (GRF) PolyU 5245/09E.

\appendix

\subsection{Proof of the NP-hardness of RMRP}
\label{sec:rmrp_np}

\begin{definition}
We call the special case of RMRP problem, where the robustness index $\rho =
0$, the MRP problem.
\end{definition}
\vspace{0.01in}

\begin{lemma} \label{NP_RMRP1}
MRP is NP-hard.
\end{lemma}

\begin{proof}
Since $\rho = 0 \Rightarrow \Gamma_k \equiv 0$ in MRP, the \emph{protection
function} of \eqref{scheduling_dnorm} is null, which means only primary paths
are considered in \eqref{scheduling}.  In MRP, $r_i\tau_{ik}$ represents the
percentage of relay $k$'s capacity occupied by logical link $i$'s primary path,
if $i$ choose to route its primary path via $k$.  If we treat a bin as a
relay's capacity, and treat the volume of an item as the percentage of
relay capacity occupied by an logical link, the \emph{Bin-Packing} problem can
be reduced to MRP in a polynomial time.  Since \emph{Bin-Packing} is
NP-hard~\cite{cormen_algorithm}, MRP is also a NP-hard problem.
\end{proof}

As MRP is just a special case of RMRP, RMRP is henceforth harder than MRP, so
RMRP is also NP-Hard.

\subsection{Proof of the NP-hardness of RMURP}
\label{sec:rmurp_np}

\begin{definition}
We call the special case of RMURP problem, where the mmWave network adopts
robustness index of $\rho = 0$ and only has one candidate relay location,
as MURP problem.
\end{definition}
\vspace{0.01in}

\begin{lemma} \label{NP_RMURP}
MURP is NP-hard.
\end{lemma}

\begin{proof}
Since $\rho = 0 \Rightarrow \Gamma_k \equiv 0$ in MURP, the \emph{protection
function} term is null.  Also, consider only one candidate relay location
$k$ in the network, the scheduling constraint of MURP in
\eqref{scheduling_rmurp} can be rewritten as: $\sum_{i} \eta_i x_{ik} \alpha
r_i \tau_{i} \le z_k$.

Let $\alpha r_i$ represent the value of item $i$, $ \eta_i \alpha r_i \tau_{i}$
represent the weight of item $i$.  the 0-1 knapsack problem can be reduced to
MURP in a polynomial time.  Since the \emph{Knapsack} problem is
NP-hard~\cite{cormen_algorithm}, MURP is NP-hard.
    \end{proof}

As MURP is a special case of RMURP, RMURP is henceforth harder than MURP. This
proves the lemma.



\bibliographystyle{IEEEtran}
\bibliography{ref}

\begin{thebibliography}{10}
\providecommand{\url}[1]{#1}
\csname url@samestyle\endcsname
\providecommand{\newblock}{\relax}
\providecommand{\bibinfo}[2]{#2}
\providecommand{\BIBentrySTDinterwordspacing}{\spaceskip=0pt\relax}
\providecommand{\BIBentryALTinterwordstretchfactor}{4}
\providecommand{\BIBentryALTinterwordspacing}{\spaceskip=\fontdimen2\font plus
\BIBentryALTinterwordstretchfactor\fontdimen3\font minus
  \fontdimen4\font\relax}
\providecommand{\BIBforeignlanguage}[2]{{%
\expandafter\ifx\csname l@#1\endcsname\relax
\typeout{** WARNING: IEEEtran.bst: No hyphenation pattern has been}%
\typeout{** loaded for the language `#1'. Using the pattern for}%
\typeout{** the default language instead.}%
\else
\language=\csname l@#1\endcsname
\fi
#2}}
\providecommand{\BIBdecl}{\relax}
\BIBdecl

\bibitem{anderson04}
C.~Anderson and T.~Rappaport, ``In-building wideband partition loss
  measurements at 2.5 and 60{GHz},'' \emph{IEEE Transactions on Wireless
  Communication}, vol.~3, no.~3, pp. 922--928, May 2004.

\bibitem{daniels10}
R.~C. Daniels, J.~N. Murdock, T.~S. Rappaport, and R.~W. Heath, ``60{GHz}
  wireless: Up close and personal,'' \emph{IEEE Microwave Magazine}, vol.~11,
  no.~7, pp. 44--50, Dec. 2010.

\bibitem{flyway2011}
D.~Halperin, S.~Kandula, J.~Padhye, P.~Bahl, and D.~Wetherall, ``Augmenting
  data center networks with multi-gigabit wireless links,'' in \emph{ACM
  SIGCOMM}, Aug. 2011.

\bibitem{yiu09}
C.~Yiu and S.~Singh, ``Empirical capacity of {mmWave WLANs},'' \emph{IEEE
  Journal Selected Area in Communications}, vol.~27, no.~8, pp. 1479--1487,
  Oct. 2009.

\bibitem{singh09}
S.~Singh, R.~Mudumbai, and U.~Madhow, ``Medium access control for 60{GHz}
  outdoor mesh networks with highly directional links,'' in \emph{IEEE
  INFOCOM}, Apr. 2009.

\bibitem{singh10}
------, ``Distributed coordination with deaf neighbors: efficient medium access
  for 60{GHz} mesh networks,'' in \emph{IEEE INFOCOM}, May 2010.

\bibitem{singh07}
S.~Singh, F.~Ziliotto, U.~Madhow, E.~M. Belding-Royer, and M.~J.~W. Rodwell,
  ``Millimeter wave {WPAN}: Cross-layer modeling and multi-hop architecture,''
  in \emph{IEEE INFOCOM}, May. 2007.

\bibitem{yiu10}
C.~Yiu and S.~Singh, ``Link selection for point-to-point 60{GHz} networks,'' in
  \emph{IEEE ICC}, May 2010.

\bibitem{Jian-twc11}
J.~Qiao, L.~X. Cai, X.~Shen, and J.~W. Mark, ``Enabling multi-hop concurrent
  transmissions in 60 ghz wireless personal area networks,'' \emph{IEEE
  Transactions on Wireless Communications}, vol.~10, no.~11, pp. 3824--3833,
  Nov. 2011.

\bibitem{Alberto}
\BIBentryALTinterwordspacing
A.~Valdes-Garcia, ``Millimeter-wave communications using a reflector,'' Patent
  US 2012/0\,206\,299 A1, 08 16, 2012. [Online]. Available:
  \url{http://www.google.com/patents/US20120206299.}
\BIBentrySTDinterwordspacing

\bibitem{Kenneth}
\BIBentryALTinterwordspacing
K.~W. Brown, ``Directed energy beam virtual fence,'' Patent US 2009/0\,256\,706
  A1, 10 15, 2009. [Online]. Available:
  \url{http://www.google.com/patents/US20090256706.}
\BIBentrySTDinterwordspacing

\bibitem{xia2012}
X.~Zhou, Z.~Zhang, Y.~Zhu, Y.~Li, S.~Kumar, A.~Vahdat, B.~Y. Zhao, and
  H.~Zheng, ``Mirror mirror on the ceiling: Flexible wireless links for data
  centers,'' in \emph{ACM SIGCOMM}, Aug. 2012.

\bibitem{yang091}
K.~Yang, J.~Huang, Y.~Wu, X.~Wang, and M.~Chiang, ``Distributed robust
  optimization part {I}: Framework and example,'' in \emph{Technical Report},
  Princeton University, Jan. 2009.

\bibitem{sum-gc09}
C.-S. Sum, Z.~Lan, and etc, ``A multi-gbps {Millimeter-wave WPAN} system based
  on {STDMA} with heuristic scheduling,'' in \emph{IEEE Globecom}, Nov. 2009.

\bibitem{cai-wcnc07}
L.~X. Cai, L.~Cai, X.~Shen, and J.~W. Mark, ``Efficient resource management for
  {mmWave WPANs},'' in \emph{IEEE WCNC}, Mar. 2007.

\bibitem{cai-gc07}
------, ``Spatial multiplexing capacity analysis of mm{W}ave {WPAN}s with
  directional antennae,'' in \emph{IEEE GLOBECOM}, Nov. 2007.

\bibitem{cai-twc10}
------, ``{Rex}: A randomized exclusive region based scheduling scheme for
  {mmWave WPANs} with directional antenna,'' \emph{IEEE Transactions on
  Wireless Communications}, vol.~9, no.~1, pp. 113--121, Jan. 2010.

\bibitem{singh-jsac09}
S.~Singh, F.~Ziliotto, U.~Madhow, E.~M. Belding, and M.~Rodwell, ``Blockage and
  directivity in 60 {GHz} wireless personal area networks: From cross-layer
  model to multihop mac design,'' \emph{IEEE Journal Selected in
  Communications}, vol.~19, no.~5, pp. 1513--1527, Oct. 2009.

\bibitem{gong-wcnc10}
M.~X. Gong, R.~Stacey, D.~Akhmetov, and S.~Mao, ``A directional {CSMA/CA}
  protocol for mm{W}ave wireless {PAN}s,'' in \emph{IEEE WCNC}, Apr. 2010.

\bibitem{gong-gc10}
M.~X. Gong, D.~Akhmetov, R.~Want, and S.~Mao, ``Directional {CSMA/CA} protocol
  with spatial reuse for {mmWave} wireless networks,'' in \emph{IEEE GLOBECOM},
  Dec. 2010.

\bibitem{gong-icc11}
------, ``Multi-user operation in mm{W}ave wireless networks,'' in \emph{IEEE
  ICC}, Jun. 2011.

\bibitem{wang10_vtc}
Z.~Lan and J.~W. etc, ``Directional relay with spatial time slot scheduling for
  {mmWave WPAN} systems,'' in \emph{IEEE VTC Spring}, May 2010.

\bibitem{Zhou2009}
Z.~Lan, C.~sean Sum, J.~Wang, T.~Baykas, J.~Gao, H.~Nakase, H.~Harada, and
  S.~Kato, ``Deflect routing for throughput improvement in muti-hop
  millimeter-wave wpan system,'' in \emph{IEEE WCNC}, Apr. 2009.

\bibitem{Partha2010}
P.~Dutta, V.~Mhatre, D.~Panigrahi, and R.~Rastogi, ``Joint routing and
  scheduling in multi-hop wireless networks with directional antennas,'' in
  \emph{IEEE Infocom}, Mar. 2010.

\bibitem{di2010}
D.~Li, C.~Yin, and C.~Chen, ``A selection region based routing protocol for
  random mobile ad hoc networks with directional antennas,'' in \emph{IEEE
  Globecom}, Dec. 2010.

\bibitem{zheng-rtas12}
G.~Zheng, C.~Hua, K.~Vu, R.~Zheng, and Q.~Wang, ``Robust reflector placement in
  60{GHz} {mmWave} wireless personal area networks,'' in \emph{IEEE RTAS,
  Work-in-Progress}, Apr. 2012.

\bibitem{Bai_mobility}
F.~Bai and A.~Helm, \emph{A Survey of Mobility Modeling and Analysis in
  Wireless Adhoc Networks}.\hskip 1em plus 0.5em minus 0.4em\relax Springer,
  Oct. 2006.

\bibitem{cplex}
\BIBentryALTinterwordspacing
I.~ILOG. {CPLEX} optimization studio for academics. [Online]. Available:
  \url{http://www-01.ibm.com/software/websphere/products/optimization
  /academic-initiative/.}
\BIBentrySTDinterwordspacing

\bibitem{li2006}
D.~Li and X.~Sun, \emph{Nonlinear Integer Programming}.\hskip 1em plus 0.5em
  minus 0.4em\relax Springer, 2006.

\bibitem{hua12_tmc}
C.~Hua and R.~Zheng, ``Robust topology engineering in multiradio multichannel
  wireless networks,'' \emph{IEEE Transactions on Mobile Computing}, vol.~11,
  no.~3, pp. 492--503, Mar. 2012.

\bibitem{cormen_algorithm}
T.~H. Cormen, \emph{Introduction to Algorithms}.\hskip 1em plus 0.5em minus
  0.4em\relax The MIT Press, Jul. 2009.

\end{thebibliography}

\end{document}